%% file: Manuscript.tex
\documentclass[a4paper,fleqn]{cas-sc}

\usepackage[numbers,sort&compress]{natbib}
\usepackage{bm}
\usepackage{tabularx}
\makeatletter
\providecommand{\insert@pcolumn}{\insert@column}
\makeatother
\ExplSyntaxOn
\RenewDocumentCommand \printorcid { }
{
  \seq_if_empty:NF \g_stm_orcid_seq
    {
      \group_begin:
        \tex_let:D \thefootnote \relax \footnotetext
        {
          \raggedright
          \textsc{orcid}(s):\c_space_token
          \seq_use:Nn \g_stm_orcid_seq { ;~ }
        }
      \group_end:
    }
}
\cs_gset:Npn \__first_footerline:
{
  \group_begin:
  \small
  \sffamily
  \ifnum\theblind>0\relax
  \else
    \__short_authors: :~
  \fi
  { \rmfamily \itshape Preprint }
  \group_end:
}
\ExplSyntaxOff

\usepackage{placeins,needspace}
\usepackage{enumitem}
\usepackage[mathlines]{lineno}
\graphicspath{{figures/}}
\input{figure_annotations.tex}

\begin{document}
\hypersetup{allcolors=black}
\raggedbottom
\emergencystretch=3em
\setlength{\linenumbersep}{10pt}
\setcounter{topnumber}{4}
\setcounter{bottomnumber}{3}
\setcounter{totalnumber}{6}
\renewcommand{\topfraction}{0.92}
\renewcommand{\bottomfraction}{0.85}
\renewcommand{\textfraction}{0.06}
\renewcommand{\floatpagefraction}{0.76}
\setlength{\textfloatsep}{9pt plus 2pt minus 2pt}
\setlength{\floatsep}{9pt plus 2pt minus 2pt}
\setlength{\intextsep}{9pt plus 2pt minus 2pt}
\makeatletter
\setlength{\@fptop}{0pt plus 1fil}
\setlength{\@fpsep}{14pt plus 0pt minus 2pt}
\setlength{\@fpbot}{0pt plus 1fil}
\makeatother

\shorttitle{}
\shortauthors{G.~Dong et~al.}

\title[mode=title]{Three-Dimensional Wake Reconstruction and Cross-Scale Transfer for Floating Offshore Wind Turbines via Physics-Informed Fourier Neural Operators}

\author[aff1,aff2,aff3]{Guodan Dong}[orcid=0000-0001-8488-8294]
\author[aff1,aff2]{Qi Qian}
\author[aff1,aff2]{Feifei Xue}
\author[aff1,aff2]{Jianhua Qin}[orcid=0000-0002-9932-400X]
\cormark[1]
\ead{J.Qin@hhu.edu.cn}
\author[aff1,aff2,aff3]{Chang Xu}
\author[aff4]{Wenzhong Shen}
\cortext[1]{Corresponding author at: College of Renewable Energy, Hohai University, Changzhou, 213200, China.}

\affiliation[aff1]{
  organization={College of Renewable Energy, Hohai University},
  city={Changzhou},
  postcode={213200},
  country={China}
}

\affiliation[aff2]{
  organization={National Technology Innovation Center for Wind Power, Hohai University},
  city={Changzhou},
  postcode={213200},
  country={China}
}

\affiliation[aff3]{
  organization={College of Water Conservancy and Hydropower Engineering, Hohai University},
  city={Nanjing},
  postcode={210098},
  country={China}
}

\affiliation[aff4]{
  organization={College of Electrical Energy and Power Engineering, Yangzhou University},
  city={Yangzhou},
  postcode={225127},
  country={China}
}

\begin{abstract}
Coupled surge--pitch motions drive unsteady vortex interactions and dynamic wake meandering behind floating offshore wind turbines, complicating three-dimensional wake reconstruction from limited measurements and transfer across turbine scales. Large-eddy simulations coupled with an actuator line model generate fixed and surge--pitch wakes at motion-based Strouhal numbers $St=fD/U_\infty$ of 0.2--0.6 (where $f$, $D$, and $U_\infty$ are the motion frequency, rotor diameter, and inflow velocity) for the National Renewable Energy Laboratory 5~MW ($D=126$~m) and International Energy Agency 22~MW ($D=284$~m) turbines. A physics-informed Fourier neural operator framework combines four-dimensional spectral convolutions with incompressible continuity and rotor-forced momentum-residual regularization to reconstruct three-dimensional unsteady wakes from two observation planes. Trained on the 5~MW fixed and $St=0.2$--$0.5$ cases, the framework recovers near-wake double-peak deficits, motion-driven wake expansion, and coherent vortices through $St=0.6$ (mean wake-center and half-width errors within 1.3\% and 1.0\% of $D$), while 10\% planar sampling preserves near-dense accuracy (relative $L_2$ error 0.0689 versus 0.0684). Transfer learning is then applied and evaluated from the smaller 5~MW turbine to the larger 22~MW turbine. Relative to zero-shot inference, transfer learning lowers the three-dimensional streamwise-velocity error of the 22~MW turbine at $St=0.6$ by 36\% (0.1066 to 0.0682), halves scratch-training error with one fixed case (0.0835 versus 0.1693), and restores wake-center tracking and motion-locked harmonics ($St$--$3St$). Large-scale motion-induced dynamics are successfully transferred across turbine scales, whereas fine-scale vortex breakdown remains damped. The framework shows potential for three-dimensional wake reconstruction and downstream-inflow assessment in floating wind farms.
\end{abstract}


\maketitle


\section{Introduction}
\label{sec:introduction}

Global offshore wind reached 92.5~GW of cumulative capacity by the end of 2025, with 9.3~GW of new installations, turbines averaging over 10~MW, and cumulative capacity forecast to exceed 420~GW by 2035~\cite{gwec2026offshore}. In deeper waters, floating offshore wind turbines (FOWTs) undergo wave- and wind-induced motions that alter rotor kinematics and excite unsteady three-dimensional (3D) wakes that govern downstream power losses and fatigue loads, which is critical for wind-farm layout design, active wake control, and structural reliability~\cite{porteagel2020review}.

Wind-turbine wakes are 3D multiscale turbulent flows shaped by blade-shed vortices, nacelle and tower wakes, shear-layer instability, and ambient entrainment. Helical tip and root vortices interact with hub structures in the near wake to establish the momentum deficit and turbulence production~\cite{dong2023bladewakes,messmer2025motionstructures}, while shear-layer roll-up and turbulent mixing drive far-wake recovery~\cite{bourhis2025fstct,dong2026tandem}. For floating rotors, platform oscillations inject low-frequency forcing directly into this turbulent wake scenario, as shown in simulations of side-to-side motion~\cite{li2022fowtmeandering} and wind-tunnel tests under multi-degree-of-freedom motions~\cite{fontanella2025opendataset}. During coupled surge--pitch excitation, streamwise surge varies rotor-normal inflow while pitching tilts the rotor disk, modulating thrust, vortex spacing, and unsteady downstream wake advection. This periodic forcing is parameterized by the motion-based Strouhal number $St = fD/U_{\infty}$, where $f$ is turbine oscillation frequency, $D$ is rotor diameter, and $U_{\infty}$ is inflow velocity. Although fixed 5~MW, 15~MW, and 22~MW turbines exhibit similar normalized far-wake statistics~\cite{parinam2026scaling}, unsteady rotor loads under surge vary markedly across designs~\cite{schulz2026surge}, and recent surge--pitch simulations confirm scale-dependent wake deformation~\cite{dong2027wake}, precluding fixed-rotor geometric scaling from 5~MW to 22~MW units. Fast, physics-consistent 3D unsteady wake modeling frameworks that adapt across rotor scales from limited measurements are thus needed for floating wind turbines.

Analytical wake models capture motion-dependent mean deficits cheaply~\cite{he2026pitchwake}, but omit 3D coherent vortices and harmonic phase lags required for load analysis. High-fidelity large-eddy simulation (LES) coupled with an actuator line model (ALM) resolves energy-containing wake eddies and blade-level induction, yet running parametric sweeps across broad envelopes of platform motion frequencies and rotor scales remains computationally expensive. Conversely, wind-tunnel experiments and field lidar or particle image velocimetry (PIV) campaigns typically observe only discrete lines or sparse two-dimensional (2D) planes rather than complete 3D volumetric time series. Data-driven methods reconstruct fine-scale flows from coarse fields~\cite{fukami2019superresolution}, sparse sensors~\cite{erichson2020shallow}, and visualizations~\cite{raissi2020hfm}. However, wake meandering is a large-scale motion of the entire wake that intensifies under floating six-degree-of-freedom motions~\cite{li2022fowtmeandering,fontanella2025opendataset}, coupling core excursions with helical vortices and making 3D wake modeling from sparse planar data challenging.

Physics-informed neural networks (PINNs)~\cite{raissi2019pinn} assimilate sparse data using PDE constraints for 2D lidar wake retrieval~\cite{wang2024wakepinnlidar,zhang2024virtualLidarPinn}, 3D wind reconstruction~\cite{chen2025multiscalepinn}, turbulent wakes~\cite{gafoor2025pinnwake}, time-partitioned inference~\cite{luo2026sparsewake}, and multifidelity transfer~\cite{wang2026tlpinn}. However, while PINNs preserve physics, per-case 4D differentiation makes multi-regime 3D prediction slow. Recently, neural operators (NOs)~\cite{chen1995operator,kovachki2023neuraloperator} have accelerated parametric inference via DeepONet~\cite{lu2021deeponet}, graph kernels~\cite{li2020graphoperator}, and Fourier neural operators (FNOs)~\cite{li2021fno}, while PDE residuals yield physics-informed DeepONets~\cite{wang2021pidon} and neural operators (PINOs)~\cite{li2024pino}. Recent operator models reconstruct channel turbulence~\cite{wang2024iufno}, predict dynamic wakes~\cite{zhang2025stfno}, resolve yawed far wakes~\cite{lee2025cfno}, emulate wind-farm LES~\cite{ahn2026surrogate}, capture FOWT harmonics~\cite{DONG2026143093}, and infer 3D wakes from coarse volumes~\cite{fang2027latentwake}. Yet purely data-driven NOs demand dense 3D supervision and can violate conservation under unseen forcing or sparse inputs. Physics-informed Fourier neural operators (PIFNOs) unite rapid space--time spectral operator learning with governing PDE residual regularization to enable physically consistent inference, yet reconstructing 3D unsteady floating turbine wakes directly from sparse orthogonal 2D planes and systematically transferring operators across distinct turbine scales remains unexplored.

Three key research gaps motivate this study. First, rapid and physically consistent 3D unsteady wake reconstruction directly from two orthogonal 2D observation planes (horizontal hub-height and vertical centerline planes), or sparse random subsets thereof, has not yet been demonstrated across out-of-training platform motion frequencies. Second, scaling from a small-scale 5~MW turbine to a large-scale 22~MW machine alters rotor diameter, hub height, blade geometry, and tip-speed ratio (TSR) concurrently, leaving cross-scale transfer efficiency relative to zero-shot scaling and scratch training unquantified. Third, global $L_2$ norms and time-averaged deficit profiles mask the damping of instantaneous wake meandering, harmonics, 3D vortices, and turbulent features, necessitating coordinated time-averaged, instantaneous, spectral, and vortical evaluation.

To bridge these gaps, we build an LES--ALM database covering the National Renewable Energy Laboratory (NREL) 5~MW ($D = 126$~m) and International Energy Agency (IEA) 22~MW ($D = 284$~m) turbines across fixed and surge--pitch cases ($St = 0.2$--$0.6$), and formulate a PIFNO framework trained on the NREL 5~MW fixed and $St = 0.2$--$0.5$ cases with evaluation at $St = 0.6$, which is then transferred to the IEA 22~MW turbine scenarios. Our main contributions are:
\begin{enumerate}[leftmargin=*,nosep]
\item \emph{Proposed a physics-informed Fourier neural operator (PIFNO) framework for 3D unsteady wake reconstruction:} We embed incompressible continuity and a learnable rotor-localized momentum source within space--time spectral operator learning to reconstruct 3D unsteady multiscale wakes directly from two orthogonal 2D planes. Trained solely on NREL 5~MW fixed and $St = 0.2$--$0.5$ cases, the framework generalizes beyond the training cases to $St = 0.6$ with a volumetric streamwise-velocity relative $L_2$ error of 0.0684 while keeping mean wake-center and half-width errors within $0.013D$ and $0.010D$.
\item \emph{Quantified sparse planar observability limits for 3D wake reconstruction:} We evaluate random planar sampling coverages from 2\% to 50\% under out-of-training cases ($St = 0.6$) across both turbines. Retaining only 10\% of planar observations preserves near-dense 3D volumetric accuracy, yielding streamwise-velocity relative $L_2$ errors of 0.0689 compared with 0.0684 for dense sampling on the NREL 5~MW turbine and 0.0716 compared with 0.0682 on the transferred IEA 22~MW turbine.
\item \emph{Developed cross-scale transfer learning across utility-scale wind turbines:} We establish a cross-scale transfer learning strategy from the small-scale NREL 5~MW turbine to the large-scale IEA 22~MW turbine, comparing zero-shot inference, one-shot, three-shot, and full transfer against training from scratch. At 22~MW $St = 0.6$, full transfer lowers 3D volumetric error by 36\% relative to zero-shot scaling (from 0.1066 to 0.0682), and one-shot transfer halves scratch-training error (from 0.1693 to 0.0835).
\item \emph{Conducted multiscale turbulent wake evaluations:} We coordinate time-averaged deficit profiles and expansion, instantaneous wake-center trajectories, pre-multiplied power spectral density (FPSD), and 3D $Q$-criterion vortices. The diagnostics decouple transferable large-scale motion-driven wake dynamics and forcing harmonics ($St$, $2St$, and $3St$) from partially damped fine-scale vortex breakdown and wake meandering.
\end{enumerate}

The remainder of this paper is organized as follows. Section~\ref{sec:method} presents the database, PIFNO formulation, and diagnostics. Section~\ref{sec:results} analyzes the reconstruction, sparsity, and transfer results, and Section~\ref{sec:conclusions} concludes the study.

\section{Methods}
\label{sec:method}

\begin{figure}[pos=htbp,align=\centering]
\centering
\includegraphics[width=0.9\textwidth,keepaspectratio]{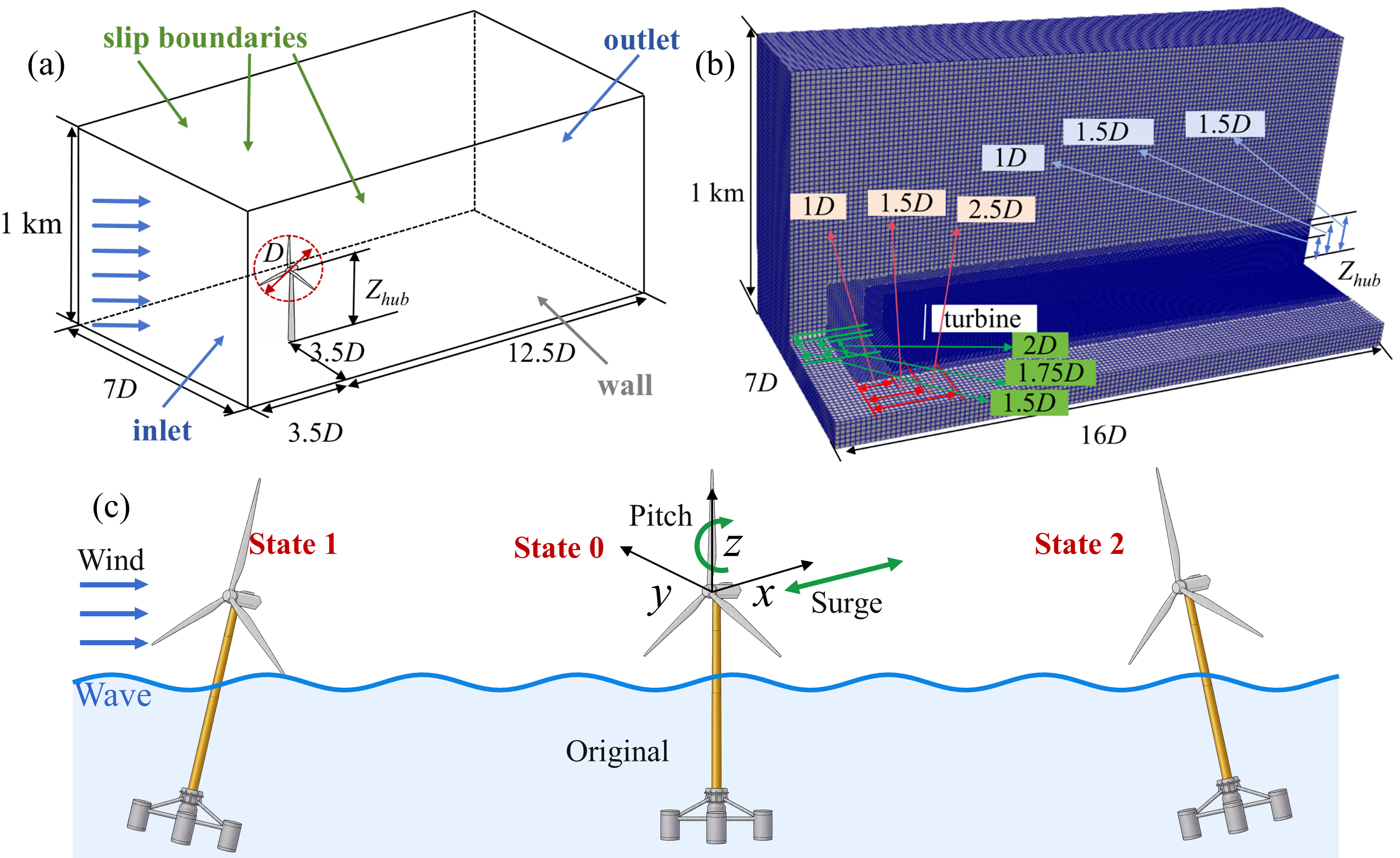}
\caption{Schematic of the LES--ALM computational setup: (a) computational domain ($16D\times 7D\times 1$~km) with turbine placed $3.5D$ downstream of the inlet and boundary conditions, (b) three-level nested mesh refinement around the rotor and wake, and (c) coupled surge--pitch motion showing the nominal configuration (State~0) and extreme displaced--pitched positions (States~1 and 2) relative to the hub-centered axes.}
\label{fig:domain_setup}
\end{figure}

This section details the LES--ALM solver, computational grid, and coupled surge--pitch wake database across two utility-scale wind turbines (Section~\ref{sec:les_setup}), formulates the nondimensional wake-reconstruction task conditioned on orthogonal planar slices (Section~\ref{sec:wake_data}), and describes the PIFNO architecture, momentum-regularized objective, sparse and cross-turbine transfer protocols, and scale-resolved turbulent wake diagnostics (Section~\ref{sec:pifno}), building on the verified numerical setups of Dong et al.~\cite{DONG2026143093,dong2027wake}. Figure~\ref{fig:domain_setup} illustrates the computational domain, nested mesh topology, and prescribed surge--pitch motions.

\subsection{LES--ALM formulation and computational setup}
\label{sec:les_setup}
High-fidelity wake reference fields are generated using incompressible LES in OpenFOAM~\cite{weller1998openfoam}, with rotating rotor aerodynamics modeled via an actuator line formulation. The solver configuration, discretization, and actuator line parameterization follow established FOWT wake studies under coupled surge--pitch oscillations~\cite{DONG2026143093,dong2027wake}, encompassing the NREL 5~MW and IEA 22~MW wake datasets. The governing Navier--Stokes equations are the spatially filtered incompressible continuity and momentum equations:
\begin{linenomath}
\postdisplaypenalty=0
\begin{gather}
\frac{\partial \tilde{u}_i}{\partial x_i}=0,
\label{eq:les_continuity}
\\
\frac{\partial \tilde{u}_i}{\partial t}
+\frac{\partial (\tilde{u}_i\tilde{u}_j)}{\partial x_j}
=-\frac{1}{\rho}\frac{\partial \tilde{p}}{\partial x_i}
+\nu\frac{\partial^2\tilde{u}_i}{\partial x_j\partial x_j}
-\frac{\partial\tau_{ij}}{\partial x_j}
+\frac{f_{{\rm WT},i}}{\rho},
\label{eq:les_momentum}
\end{gather}
\end{linenomath}
where $x_i$ ($i=1,2,3$) denote streamwise, spanwise, and vertical Cartesian coordinates, $t$ is time, repeated indices follow summation convention, $\rho$ is air density, $\nu$ is kinematic viscosity, $f_{{\rm WT},i}$ represents the turbine aerodynamic body force per unit volume, and $\tilde{u}_i$ and $\tilde{p}$ are the resolved velocity and pressure fields, respectively. Unresolved small-scale motions enter through the subgrid-scale (SGS) stress tensor $\tau_{ij}=\widetilde{u_i u_j}-\tilde{u}_i\tilde{u}_j$, modeled via the dynamic Smagorinsky closure:
\begin{linenomath}
\begin{equation}
\tau_{ij}-\frac{1}{3}\tau_{kk}\delta_{ij}
=-2\nu_{\rm sgs}\tilde{S}_{ij},\qquad
\nu_{\rm sgs}=(C_s\Delta)^2|\tilde{S}|,
\label{eq:sgs_closure}
\end{equation}
\end{linenomath}
where $\delta_{ij}$ is the Kronecker delta, $\tilde{S}_{ij}=(\partial\tilde{u}_i/\partial x_j+\partial\tilde{u}_j/\partial x_i)/2$ is the resolved strain-rate tensor, $|\tilde{S}|=(2\tilde{S}_{ij}\tilde{S}_{ij})^{1/2}$, $\Delta$ is the local filter width taken as the cube root of the computational cell volume, and the Smagorinsky coefficient $C_s$ is determined dynamically during execution (with the isotropic trace $\tfrac{1}{3}\tau_{kk}\delta_{ij}$ absorbed into the modified pressure).

Each rotating blade is discretized along its span into discrete actuator elements~\cite{turbinesFoam}, complemented by distributed body forces representing nacelle drag and tower interference. For an actuator element of spanwise length ${\rm d}r$, the local aerodynamic load vector is evaluated from blade-element momentum principles as:
\begin{linenomath}
\begin{equation}
\bm{F}=\frac{1}{2}\rho c|\bm{U}_{\rm rel}|^2
(C_L\bm{e}_L+C_D\bm{e}_D){\rm d}r,
\label{eq:alm_force}
\end{equation}
\end{linenomath}
where $c$ is the local blade chord, $C_L$ and $C_D$ are the airfoil sectional lift and drag coefficients tabulated as functions of local angle of attack, $\bm{e}_L$ and $\bm{e}_D$ are unit vectors in the lift and drag directions, and the relative inflow velocity seen by the moving blade section is $\bm{U}_{\rm rel}=\bm{U}_{\rm in}-\bm{V}_{\rm motion}-\bm{V}_{\rm rotation}$, where $\bm{U}_{\rm in}$, $\bm{V}_{\rm motion}$, and $\bm{V}_{\rm rotation}$ denote the local fluid velocity sampled at the actuator point, the rigid-body velocity induced by turbine motion, and the tangential velocity due to rotor rotation, respectively. Equal and opposite aerodynamic reaction forces are distributed onto surrounding Eulerian mesh cells using an isotropic 3D Gaussian regularization kernel~\cite{matineztossas2017smoothing}:
\begin{linenomath}
\begin{equation}
\bm{f}_{\rm WT}(\bm{x},t)=-\sum_n\bm{F}_n(t)\eta(d_n),\qquad
\eta(d)=\frac{\exp[-(d/\varepsilon)^2]}{\varepsilon^3\pi^{3/2}},
\label{eq:gaussian_kernel}
\end{equation}
\end{linenomath}
where $d_n$ is the Euclidean distance from cell center $\bm{x}$ to the $n$-th actuator node and $\varepsilon=\max(0.25c,C_Dc/2,2\Delta)$ suppresses numerical oscillations near the blade. Coupled surge--pitch platform motion therefore perturbs the unsteady wake through two tightly coupled pathways: displacing actuator trajectories across the stationary mesh and altering $\bm{V}_{\rm motion}$ inside $\bm{U}_{\rm rel}$.

As shown in Fig.~\ref{fig:domain_setup}(a), the computational domain extends $16D\times 7D\times 1$~km in the streamwise, spanwise, and vertical directions, with the turbine positioned on the spanwise centerline $3.5D$ downstream of the inlet boundary to provide $12.5D$ of wake development. Three nested Cartesian refinement zones progressively refine the background $D/8$ resolution to a uniform $D/64$ grid ($\approx 2\times10^7$ cells total, Fig.~\ref{fig:domain_setup}(b)), with the finest region covering $x/D\in[-1,12.5]$, $|y/D|\le 1.5$, and $0\le z\le z_{\rm hub}+D$ to resolve near-wake tip vortices, hub separation, and far-wake meandering. Boundary conditions specify uniform streamwise inflow ($U_\infty=11.4$~m/s) with zero normal pressure gradient at the inlet, zero velocity gradient with fixed static pressure at the outlet, a smooth no-slip wall along the bottom plane ($z=0$), and free-slip conditions on lateral and top boundaries.

Table~\ref{tab:turbines} summarizes the geometric dimensions and operating parameters of the NREL 5~MW~\cite{jonkman2009nrel5mw} and IEA 22~MW~\cite{zahle2024iea22mw} reference wind turbines. Six cases are simulated for each wind turbine: a fixed-bottom baseline ($A_s=A_p=0$) and five coupled surge--pitch cases spanning $St=0.2,0.3,0.4,0.5,0.6$. Prescribed streamwise surge displacement $x_s(t)$ and hub-centered pitch angle $\theta_p(t)$ (Fig.~\ref{fig:domain_setup}(c)) follow synchronized sinusoidal waveforms, $x_s(t)=A_s\sin(2\pi ft)$ and $\theta_p(t)=A_p\sin(2\pi ft)$, with fixed normalized surge amplitude $A_s=0.04D$ and pitch amplitude $A_p=5^{\circ}$ across all oscillating cases. Both turbines operate at their rated inflow speed $U_\infty=11.4$~m/s with tip-speed ratios $\lambda=\Omega D/(2U_\infty)$ (where $\Omega$ is the turbine angular speed) of 8.0 for the NREL 5~MW turbine and 9.1 for the IEA 22~MW turbine. Spatial grid-convergence verification at $D/64$ and blade-load validation against benchmark data are established in Dong et al.~\cite{DONG2026143093,dong2027wake} and are not repeated here.

\begin{table}[pos=htbp]
\centering
\caption{Main parameters and operating conditions of the NREL 5~MW and IEA 22~MW reference turbines.}
\label{tab:turbines}
\small
\setlength{\tabcolsep}{16pt}
\renewcommand{\arraystretch}{1.12}
\begin{tabular}{@{}lcc@{}}
\toprule
Parameter & NREL 5~MW turbine & IEA 22~MW turbine \\
\midrule
Rated power (MW) & 5 & 22 \\
Hub height (m) & 90 & 170 \\
Rotor diameter (m) & 126 & 284 \\
Hub diameter (m) & 3.0 & 8.4 \\
Inflow velocity (m/s) & 11.4 & 11.4 \\
Tip-speed ratio & 8.0 & 9.1 \\
\bottomrule
\end{tabular}
\end{table}

\subsection{Wake data and reconstruction problem}
\label{sec:wake_data}
\begin{figure}[pos=htbp,align=\centering]
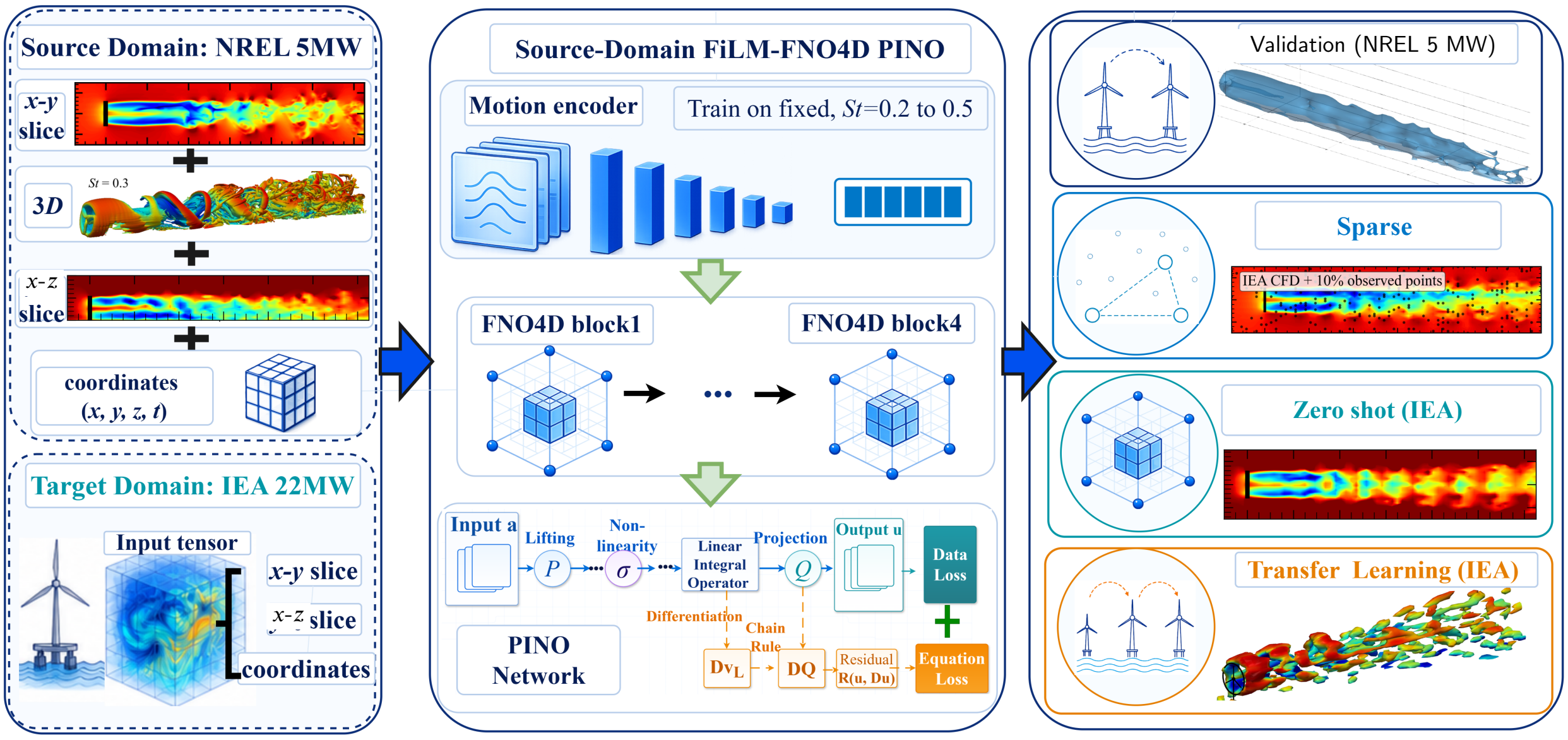

\centering
\workflowgraphic
\caption{Workflow of the wake reconstruction and cross-turbine transfer study. A physics-informed Fourier neural operator (PIFNO) is trained on NREL 5~MW cases (fixed and $St=0.2$--$0.5$) and evaluated on NREL $St=0.6$ validation, sparse planar reconstruction, zero-shot IEA 22~MW generalization, and transfer learning.}
\label{fig:workflow}
\end{figure}

To eliminate initial numerical transients, the first 250~s of simulation are discarded for the NREL 5~MW turbine and the first 500~s for the larger IEA 22~MW turbine. Both spin-up intervals and the subsequent data-acquisition windows substantially exceed one full convective flow-through time ($16D/U_\infty$) and span 57.6 and 58.1 rotor revolutions, respectively, yielding statistically stationary periodic wake states. All spatial and temporal variables are nondimensionalized by rotor diameter $D$ and freestream velocity $U_\infty$ relative to the turbine center at hub height: $x_*=x/D$, $y_*=y/D$, $z_*=(z-z_{\rm hub})/D$, and $t_*=(t-t_{\rm start})U_\infty/D$, where $t_{\rm start}$ denotes the start of the retained window and subscript $*$ identifies nondimensional physical quantities. Under these hub-centered coordinates, both turbine datasets are mapped onto a unified regular reconstruction box spanning $x_*\in[-1.5,12.5]$, $y_*\in[-1.5,1.5]$, and $z_*\in[-1,2.5]$. Because the nondimensional hub-height ratios differ between turbines ($z_{\rm hub}/D=0.71$ for NREL 5~MW and $0.60$ for IEA 22~MW), grid nodes falling below the bottom boundary ($z<0$, corresponding to $z_*<-z_{\rm hub}/D$) lie outside the fluid domain and are excluded by a binary fluid-domain mask during error evaluation. As illustrated in Fig.~\ref{fig:workflow}, planar input observations are extracted via nearest-grid projection onto two orthogonal planes, namely, the horizontal hub-height plane ($z=z_{\rm hub}$) and the vertical centerline plane ($y=0$), while available 3D volumetric snapshots provide ground-truth supervision during training.

The nondimensional flow state is defined as $\bm{q}_*=(u_*,v_*,w_*,p_*)$, where velocity components are normalized by $U_\infty$ and kinematic pressure is scaled as $p_*=(p/\rho)/U_\infty^2$. To balance velocity and pressure magnitudes during operator learning, each channel $c\in\{u,v,w,p\}$ is standardized according to $\bm{q}_n=(\bm{q}_*-\bm{\mu})/\bm{s}$, where $s_c=\sigma_c+10^{-6}$ is stabilized from the channel mean $\mu_c$ and standard deviation $\sigma_c$ (with dataset splitting detailed in Section~\ref{sec:training_protocol}). Rescaled space--time coordinates, initial states, orthogonal planar slices, and global motion descriptors are concatenated along the channel dimension into the 20-channel dense input tensor $\bm{a}$:
\begin{linenomath}
\postdisplaypenalty=0
\begin{gather}
\bm{a}=[\bm{c},\bm{q}_{0,n},\bm{q}_{xy,n},\bm{q}_{xz,n},\bm{s}_o],
\label{eq:dense_input}
\\
\mathcal{G}_{\theta}:\bm{a}\mapsto\hat{\bm{q}}_n.
\label{eq:operator_map}
\end{gather}
\end{linenomath}
Here, $\bm{c}=(x_n,y_n,z_n,t_n)\in[-1,1]^4$ denotes normalized space--time coordinates (4 channels), while $\bm{q}_{0,n}$, $\bm{q}_{xy,n}$, and $\bm{q}_{xz,n}$ broadcast the four-channel initial 3D field, hub-height $xy$ slice, and centerline $xz$ slice along the temporal $t$, vertical $z$, and spanwise $y$ axes onto the full 4D reconstruction lattice ($3\times 4=12$ flow channels). Together with $\bm{s}_o$, which broadcasts the four-component motion condition across $(x,y,z)$ (4 channels), this yields $C_{\rm in}=20$ total dense input channels. The parametric operator $\mathcal{G}_{\theta}$ maps the conditioned input field $\bm{a}$ directly to the reconstructed 4D standardized state $\hat{\bm{q}}_n$~\cite{kovachki2023neuraloperator}. Whenever ground-truth 3D fields at $t_*=0$ are withheld (and in all sparse-observation runs), $\bm{q}_{0,n}$ is synthesized purely from the initial planar observations using the analytical fusion rule specified in Section~\ref{sec:training_protocol}. The four-component motion descriptor is written as:
\begin{linenomath}
\begin{equation}
\bm{s}_o=(St,St^2,\sin\psi,\cos\psi),\qquad
\psi=2\pi St\,\frac{t_*}{T_*},
\label{eq:motion_descriptor}
\end{equation}
\end{linenomath}
where $T_*$ is the nondimensional observation window length (with $St=0$ representing the fixed-rotor baseline). Retaining both linear and quadratic Strouhal terms accommodates nonlinear thrust-modulation trends across forcing regimes, while $\psi$ encodes the $St$-scaled dimensionless window phase coordinate rather than the physical platform oscillation phase $2\pi f t$ defined in Section~\ref{sec:les_setup}.

\subsection{PIFNO formulation, training, and evaluation}
\label{sec:pifno}
\subsubsection{Space--time operator}
\begin{figure}[pos=htbp,align=\centering]
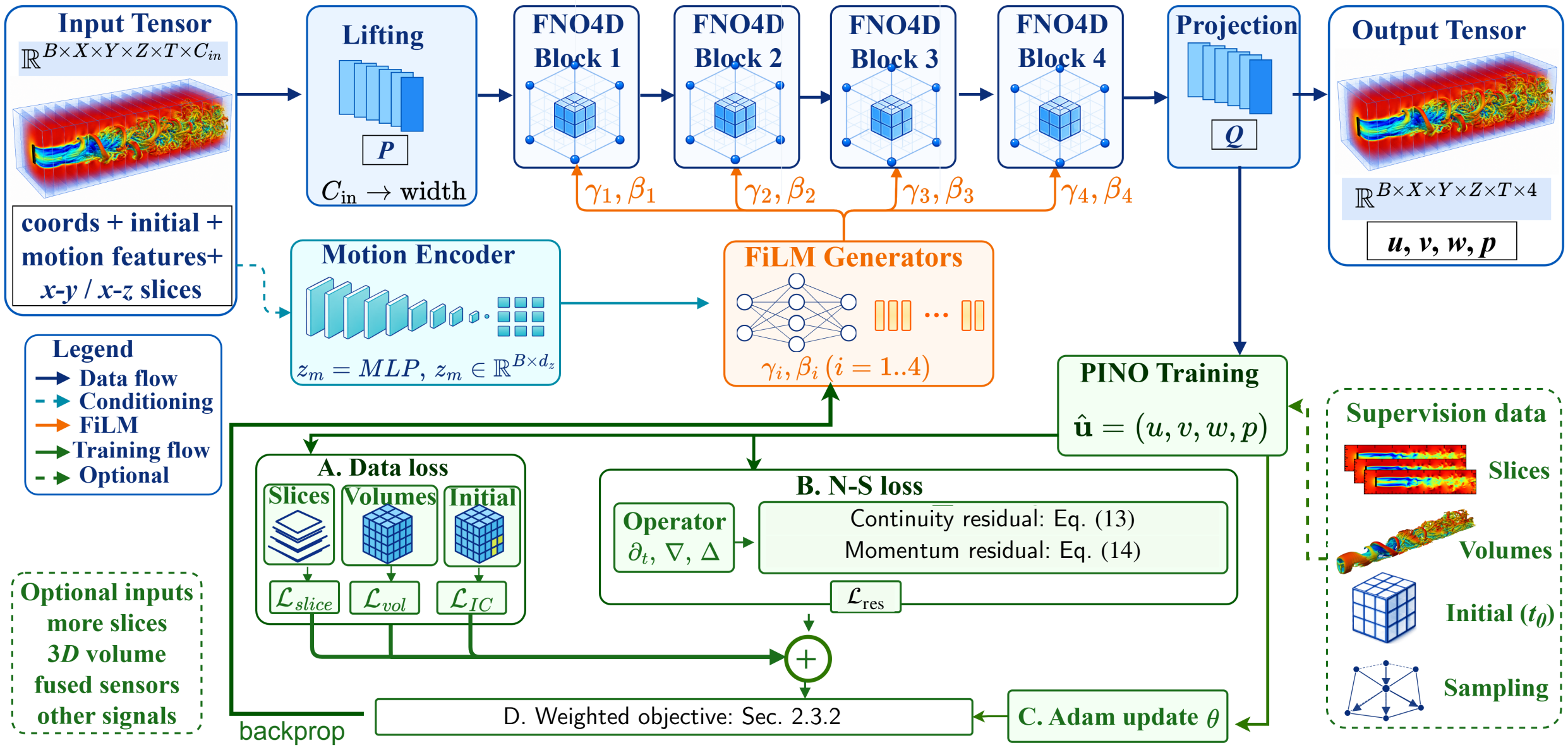

\centering
\architecturegraphic
\caption{Architecture of the PIFNO framework. Coordinates, the initial field, planar observations (or sparse samples with masks), and motion descriptors are lifted to a latent representation and passed through four Fourier blocks to reconstruct $(u,v,w,p)$. Training combines planar, volumetric, and initial-condition data losses with continuity and momentum residuals (Eqs.~\eqref{eq:div_res} and \eqref{eq:mom_res}), and loss weights, source penalties, and optimization details are given in Sections~\ref{sec:training_objective} and \ref{sec:training_protocol}.}
\label{fig:architecture}
\end{figure}

Building on physics-informed neural operators developed for turbulent flows and FOWT wakes~\cite{li2024pino,DONG2026143093}, Fig.~\ref{fig:architecture} summarizes the PIFNO architecture. A pointwise linear lifting layer first projects the $C_{\rm in}$ input channels ($C_{\rm in}=20$ for dense planar observations and $C_{\rm in}=22$ when binary sparse-sampling masks are appended) onto a 24-channel latent representation $\bm{h}^0$. The latent field then passes through four stacked 4D Fourier layers ($l=1,\dots,4$) designed to capture global advection, shear-layer coupling, and periodic wake oscillations simultaneously across space and time. Within each block, the spectral integral operator acts jointly over $(x,y,z,t)$ as:
\begin{linenomath}
\begin{equation}
\mathcal{K}_l\bm{h}^{l-1}
=\mathcal{F}^{-1}_{4D}
\left[\bm{R}_l(\bm{k})\mathcal{F}_{4D}(\bm{h}^{l-1})(\bm{k})\right],
\label{eq:fno_kernel}
\end{equation}
\end{linenomath}
where $\mathcal{F}_{4D}$ and $\mathcal{F}^{-1}_{4D}$ denote the forward and inverse 4D fast Fourier transforms, $\bm{k}=(k_x,k_y,k_z,k_t)$ indexes the retained low-frequency spectral modes with cutoff counts $(12,10,10,8)$ along $(x,y,z,t)$, and the complex tensor $\bm{R}_l(\bm{k})$ parameterizes channel mixing on retained modes while filtering high-frequency grid noise~\cite{li2021fno,kovachki2023neuraloperator}. To allow spectral and local pathways to adapt smoothly across motion Strouhal numbers and turbine regimes, feature-wise linear modulation (FiLM)~\cite{perez2018film} conditions each Fourier block via affine channel-wise scaling and shifting:
\begin{linenomath}
\begin{equation}
\bm{h}^{l}=\sigma_l\!\left[
(\bm{1}+\bm{\gamma}_l)\odot(\mathcal{K}_l\bm{h}^{l-1}+W_l\bm{h}^{l-1})
+\bm{\beta}_l\right],
\label{eq:film_block}
\end{equation}
\end{linenomath}
where $\bm{1}\in\mathbb{R}^{24}$ is an all-ones vector, $W_l$ is a local pointwise linear bypass capturing non-periodic spatial trends, and $\odot$ denotes elementwise multiplication. The affine modulation parameters $(\bm{\gamma}_l,\bm{\beta}_l)\in\mathbb{R}^{24}$ are generated by a shared two-hidden-layer conditioning network (hidden width 64) taking the space--time mean of $\bm{s}_o$, while the full time-varying $\bm{s}_o$ remains concatenated in $\bm{a}$ to preserve phase localization. The activation function $\sigma_l$ is GELU for layers $l=1,2,3$ and the identity map for the final layer $l=4$.

From $\bm{h}^4$, a two-layer pointwise projection multilayer perceptron (MLP, hidden width 160) reconstructs the low-wavenumber spectral baseline $\hat{\bm{q}}_{{\rm FNO},n}$. Because mode truncation in Eq.~\eqref{eq:fno_kernel} smooths steep near-wake velocity gradients and localized shear layers, a pointwise residual detail decoder $\mathcal{D}_{\vartheta}$ (two hidden layers of width 192, scaled by residual weight $\alpha=0.75$) refines the baseline output using both inputs and FNO predictions:
\begin{linenomath}
\begin{equation}
\hat{\bm{q}}_n=\hat{\bm{q}}_{{\rm FNO},n}
+\alpha\mathcal{D}_{\vartheta}([\bm{a},\hat{\bm{q}}_{{\rm FNO},n}]),
\label{eq:detail_head}
\end{equation}
\end{linenomath}
allowing high-wavenumber gradients and coherent wake boundaries to be sharpened locally without introducing an explicit algebraic turbulence closure.

\subsubsection{Momentum regularization and training objective}
\label{sec:training_objective}
Network training pairs supervised data misfits with PDE residual constraints evaluated directly on physical nondimensional predictions $\hat{\bm{q}}_*=\bm{s}\odot\hat{\bm{q}}_n+\bm{\mu}$ via second-order centered finite differences on the regular space--time grid~\cite{li2024pino}. Evaluating derivatives in physical nondimensional coordinates preserves consistent kinematic scaling between the NREL 5~MW and IEA 22~MW turbines. Because the rotor thrust discontinuity cannot be represented by homogeneous Navier--Stokes equations alone, we introduce a lightweight surrogate rotor-localized momentum source $\bm{f}_r$ that approximates net turbine thrust extraction and near-rotor forcing on the coarse wake lattice:
\begin{linenomath}
\begin{equation}
\bm{f}_r(\bm{x}_*,t_*)=M_r(\bm{x}_*)\bm{A}(t_*),\qquad
\bm{A}(t_*)=B\bm{\phi}_s(t_*)+\bm{b},
\label{eq:residual_force}
\end{equation}
\end{linenomath}
where the fixed spatial envelope centered on the nominal rotor disk is defined as:
\begin{linenomath}
\[
M_r(\bm{x}_*)\propto\exp\!\left[-\left(\frac{x_*}{0.18}\right)^2\right]
\left[1+\exp\!\left(\frac{\sqrt{y_*^2+z_*^2}-0.5}{0.05}\right)\right]^{-1},
\]
\end{linenomath}
which is normalized to unit peak amplitude. Unlike the moving actuator kernel $\eta(d)$ in Eq.~\eqref{eq:gaussian_kernel}, $M_r$ remains stationary around $x_*=0$ to provide a smooth support envelope for coarse-grid momentum closure. The 14-component harmonic and spline temporal basis $\bm{\phi}_s(t_*)$ (Eq.~\eqref{eq:full14_basis}, \ref{app:nrel_ablation}), trainable linear weight matrix $B\in\mathbb{R}^{3\times14}$, and bias vector $\bm{b}\in\mathbb{R}^{3}$ add only 45 learnable parameters that act exclusively inside the momentum residual penalty during training.

In these nondimensional coordinates, the continuity and momentum residuals take the form:
\begin{linenomath}
\postdisplaypenalty=0
\begin{gather}
\mathcal{C}=\nabla_*\cdot\hat{\bm{u}}_*,
\label{eq:div_res}
\\
\bm{\mathcal{M}}=\frac{\partial\hat{\bm{u}}_*}{\partial t_*}
+\hat{\bm{u}}_*\cdot\nabla_*\hat{\bm{u}}_*
+\nabla_*\hat{p}_*
-\nu_r\nabla_*^2\hat{\bm{u}}_*
-\bm{f}_r,
\label{eq:mom_res}
\end{gather}
\end{linenomath}
where $\nu_r=10^{-3}$ serves as an effective regularization diffusivity that stabilizes finite-difference Laplacians (and is distinct from both physical kinematic viscosity and LES subgrid viscosity). Averaging squared residuals across all interior space--time grid nodes, including intermediate time steps where dense 3D ground-truth snapshots are absent, yields the combined physics loss:
\begin{linenomath}
\begin{equation}
\mathcal{L}_{\rm res}=\frac{1}{4}\left[
\langle\mathcal{C}^2\rangle
+\frac{1}{3}\sum_{i=1}^3\langle\mathcal{M}_i^2\rangle\right],
\label{eq:res_loss}
\end{equation}
\end{linenomath}
and the complete composite training objective is constructed as:
\begin{linenomath}
\begin{equation}
\mathcal{L}_{\rm total}=w_{\rm slice}\mathcal{L}_{\rm slice}
+w_{\rm vol}\mathcal{L}_{\rm vol}
+w_{\rm IC}\mathcal{L}_{\rm IC}
+w_{\rm res}\mathcal{L}_{\rm res}
+w_f\mathcal{R}_f
+w_A\mathcal{R}_A,
\label{eq:total_loss}
\end{equation}
\end{linenomath}
with fixed weights $(w_{\rm slice},w_{\rm vol},w_{\rm IC},w_{\rm res})=(6,0.6,0.06,0.2)$ and $w_f=w_A=10^{-4}$. Here, $\mathcal{L}_{\rm slice}$, $\mathcal{L}_{\rm vol}$, and $\mathcal{L}_{\rm IC}$ measure standardized mean-squared errors on the two orthogonal observation planes, at available 3D volumetric supervision instants, and on the initial slice (supervised against the true initial 3D field or its dense-plane fusion), respectively. The penalty terms $\mathcal{R}_f=\langle\|\bm{f}_r\|_2^2\rangle/3$ and $\mathcal{R}_A=\langle\|\bm{A}_{j+1}-\bm{A}_j\|_2^2\rangle/3$ constrain source magnitude and step-to-step discrete variations between consecutive stored levels $j$ and $j+1$ (computed directly on consecutive steps without dividing by $\Delta t_*$).

\subsubsection{Sparse reconstruction, transfer, and model selection}
\label{sec:training_protocol}
When observations are subsampled on the two orthogonal planes, unobserved grid entries are zero-filled in standardized coordinates (equivalent to assigning each variable its channel mean) and explicitly indicated to the operator through binary observation masks $m_{xy}$ and $m_{xz}$ ($1$ at sampled nodes, $0$ elsewhere):
\begin{linenomath}
\postdisplaypenalty=0
\begin{gather}
\tilde{\bm{q}}_{xy,n}=m_{xy}\odot\bm{q}_{xy,n},\qquad
\tilde{\bm{q}}_{xz,n}=m_{xz}\odot\bm{q}_{xz,n},
\label{eq:sparse_fill}
\\
\bm{a}_{sp}=[\bm{c},\bm{q}_{0,n},\tilde{\bm{q}}_{xy,n},m_{xy},
\tilde{\bm{q}}_{xz,n},m_{xz},\bm{s}_o],
\label{eq:sparse_input}
\end{gather}
\end{linenomath}
expanding the input dimension from 20 to $C_{\rm in}=22$ channels. When ground-truth 3D initial snapshots are unavailable (and across all sparse-observation experiments), the initial streamwise velocity $u_{*,0}$ is synthesized analytically from the initial positive velocity deficits $d_{xy}=[1-u_{*,xy}(0)]_+$ and $d_{xz}=[1-u_{*,xz}(0)]_+$ via multiplicative intersection fusion:
\begin{linenomath}
\begin{equation}
d_0=\frac{d_{xy}d_{xz}}{(d_{xy,\mathrm{int}}+d_{xz,\mathrm{int}})/2+10^{-3}},\qquad
u_{*,0}=\operatorname{clip}(1-d_0,0,1.4),
\label{eq:initial_fusion}
\end{equation}
\end{linenomath}
where subscript $\mathrm{int}$ denotes values along the horizontal--vertical intersection line ($y=0,z=z_{\rm hub}$), ensuring that freestream regions with zero deficit remain undisturbed outside the wake envelope. The remaining variables ($v,w,p$) are approximated by averaging the two orthogonal slices.

To evaluate observability limits under limited measurements, sparse PIFNO observers are initialized from the converged dense NREL or fully transferred IEA weights (with weights for the two newly added mask channels initialized to zero) and fine-tuned across random, regular, and line-sampling mask patterns drawn log-uniformly between 2\% and 50\% planar coverage under full planar and available 3D training supervision. Best sparse weights are chosen at the 10\% validation coverage midpoint and subsequently frozen for testing across nested random masks spanning 2\%, 5\%, 10\%, 25\%, and 50\% coverage. In the primary NREL 5~MW training--validation split, five source cases (fixed baseline and $St=0.2$--$0.5$) form the training set while $St=0.6$ is held out for out-of-range frequency validation. For cross-turbine transfer to IEA 22~MW, pre-trained NREL weights are fine-tuned using one (fixed), three (fixed, $St=0.2,0.3$), or five cases (fixed and $St=0.2$--$0.5$, denoted full transfer) with $St=0.6$ held out, benchmarked against direct zero-shot inference and target-domain training from random initialization (scratch).

All models are optimized using AdamW (weight decay $10^{-5}$, cosine learning-rate decay, batch size 1, gradient-norm clipping at 1.0, fixed random seed 42). Dense NREL and IEA scratch runs train for 80 epochs with initial learning rate $5\times10^{-4}$, IEA transfer runs train for 60 epochs at $2\times10^{-4}$, and sparse-observation fine-tuning runs for 60 epochs at $5\times10^{-4}$.

\subsubsection{Error measures and wake diagnostics}
\label{sec:diagnostics}
Global field accuracy in physical nondimensional variables is quantified by the relative $L_2$ error $\varepsilon_{\rm rel}=\|\hat{\bm{q}}_*-\bm{q}_*\|_2/\|\bm{q}_*\|_2$, where the planar error $\varepsilon_{\rm slice}$ aggregates all four state variables $(u,v,w,p)$ across the hub-height and centerline observation planes, while the volumetric error $\varepsilon_{\rm vol}$ evaluates fluid-domain ($z\ge 0$) streamwise velocity $u_*$, which governs downstream rotor inflow power.

To move beyond integrated $L_2$ norms and examine physical wake fidelity, hub-height wake-profile diagnostics are evaluated across $2\le x/D\le 12$. At each streamwise station and instant, the reference outer-edge velocity $U_{\rm edge,*}$ is computed as the median of $u_*$ over the freestream sidebands $|y/D|>1.15$, yielding the positive velocity deficit $d_j=[U_{\rm edge,*}-u_{*,j}]_+$. First- and second-order deficit moments define the instantaneous lateral wake center $y_c/D$ and Gaussian-equivalent wake half-width $R_{1/2}/D$:
\begin{linenomath}
\begin{equation}
\frac{y_c}{D}=\frac{\sum_j(y_j/D)d_j}{\sum_jd_j},\qquad
\frac{R_{1/2}}{D}=\left[2\ln2\,
\frac{\sum_j(y_j/D-y_c/D)^2d_j}{\sum_jd_j}\right]^{1/2},
\label{eq:wake_moments}
\end{equation}
\end{linenomath}
which are time-averaged and summarized over $2\le x/D\le 12$ via spatial root-mean-square error (RMSE) and mean absolute error (MAE), where $R_{1/2}$ coincides exactly with half-maximum width strictly when the deficit profile is Gaussian. For targeted IEA meandering evaluation at $St=0.6$, instantaneous wake centers from local Gaussian curve fitting and empirical half-maximum widths are additionally extracted following spanwise Gaussian smoothing ($\sigma=0.8$ grid intervals) and scored by streamwise-averaged temporal RMSE and mean instantaneous width difference. Coherent 3D vortex rings and breakdown structures are identified using $Q$-criterion isosurfaces~\cite{hunt1988eddies}, $Q=(\|\bm{\Omega}\|_F^2-\|\bm{S}\|_F^2)/2$, where $\bm{\Omega}$ and $\bm{S}$ are the rotation-rate and strain-rate tensors and $\|\cdot\|_F$ denotes the Frobenius norm. Finally, harmonic and broadband energy recovery is evaluated through the pre-multiplied power spectral density (FPSD) of streamwise velocity fluctuations $u'$.

\Needspace{5\baselineskip}
\section{Results}
\label{sec:results}

\subsection{NREL wake reconstruction beyond the supervised frequency range}

\begin{figure}[pos=!htbp,align=\centering,abovecap=0pt]
\centering
\includegraphics[width=0.88\textwidth,keepaspectratio]{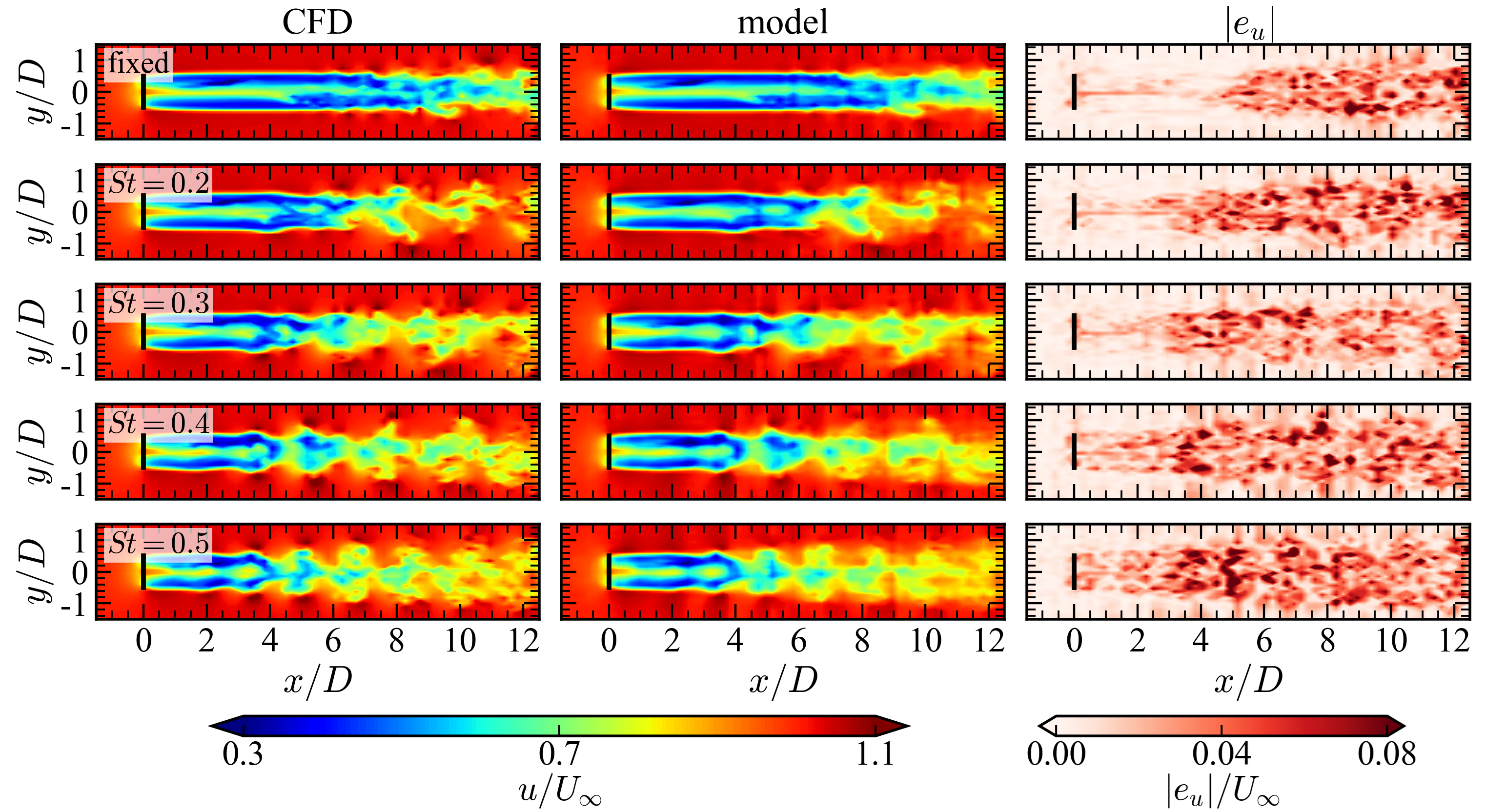}
\caption{Instantaneous hub-height streamwise velocity ($u/U_\infty$) reconstructions across the NREL 5~MW source-domain training cases (rows: fixed and $St=0.2$--$0.5$). Columns show the CFD reference, PIFNO reconstruction, and normalized error $|e_u|/U_\infty$.}
\label{fig:v6trainxy}
\end{figure}

Table~\ref{tab:v6} summarizes the relative reconstruction errors on the NREL 5~MW database when full hub-height and centerline planar observations are supplied (Section~\ref{sec:diagnostics}). Across the five supervised source-domain cases (the fixed baseline and coupled surge--pitch motions over $St=0.2$--$0.5$), planar errors remain low at 0.0225--0.0328, while 3D fluid-domain streamwise-velocity errors remain consistently within 0.0561--0.0594. As illustrated in Fig.~\ref{fig:v6trainxy}, the predicted hub-height streamwise velocity contours accurately reproduce the near-wake deficit structure, the streamwise shear-layer transition, and the periodic spatial modulation driven by coupled surge--pitch oscillations, with reconstruction errors localized primarily along steep shear-layer boundaries.

\begin{table}[pos=htbp]
\centering
\caption{Relative $L_2$ reconstruction errors for the NREL 5~MW turbine under dense planar observations. Planar errors aggregate all four channels ($u,v,w,p$), whereas 3D volume errors evaluate only the streamwise velocity $u$.}
\label{tab:v6}
\small
\setlength{\tabcolsep}{10pt}
\renewcommand{\arraystretch}{1.12}
\begin{tabular}{@{}lrrr@{}}
\toprule
Case & \shortstack{Hub-height\\plane} & \shortstack{Center\\plane} & \shortstack{3D\\volume} \\
\midrule
fixed & 0.0251 & 0.0225 & 0.0594 \\
$St=0.2$ & 0.0300 & 0.0263 & 0.0568 \\
$St=0.3$ & 0.0246 & 0.0233 & 0.0568 \\
$St=0.4$ & 0.0325 & 0.0286 & 0.0561 \\
$St=0.5$ & 0.0328 & 0.0268 & 0.0571 \\
$St=0.6$ & 0.0419 & 0.0369 & 0.0684 \\
\bottomrule
\end{tabular}
\end{table}

\begin{figure}[pos=!htbp,align=\centering,abovecap=0pt]
\centering
\includegraphics[width=0.88\textwidth,keepaspectratio]{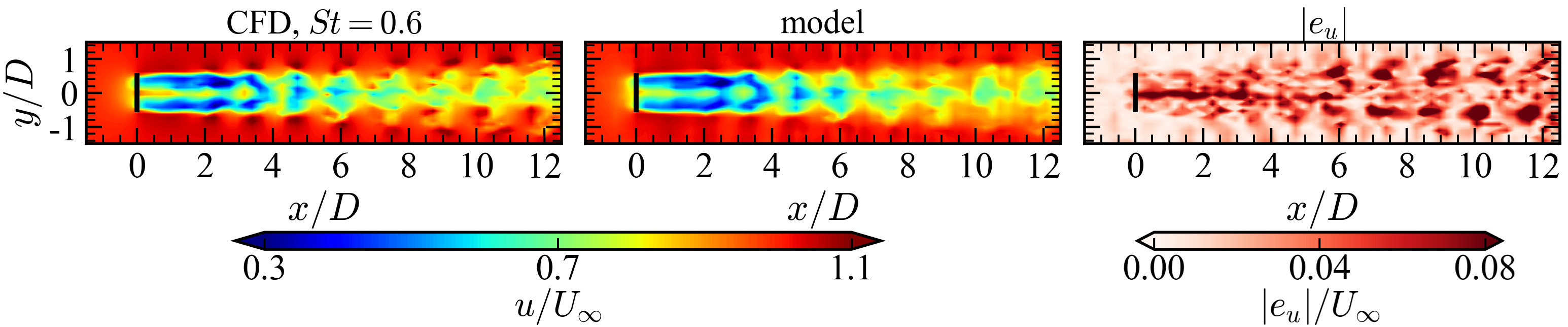}
\caption{Instantaneous hub-height streamwise velocity ($u/U_\infty$) for the NREL 5~MW validation case at $St=0.6$.}
\label{fig:v6slice}
\end{figure}

\begin{figure}[pos=!htbp,align=\centering,abovecap=0pt]
\centering
\includegraphics[width=0.82\textwidth,keepaspectratio]{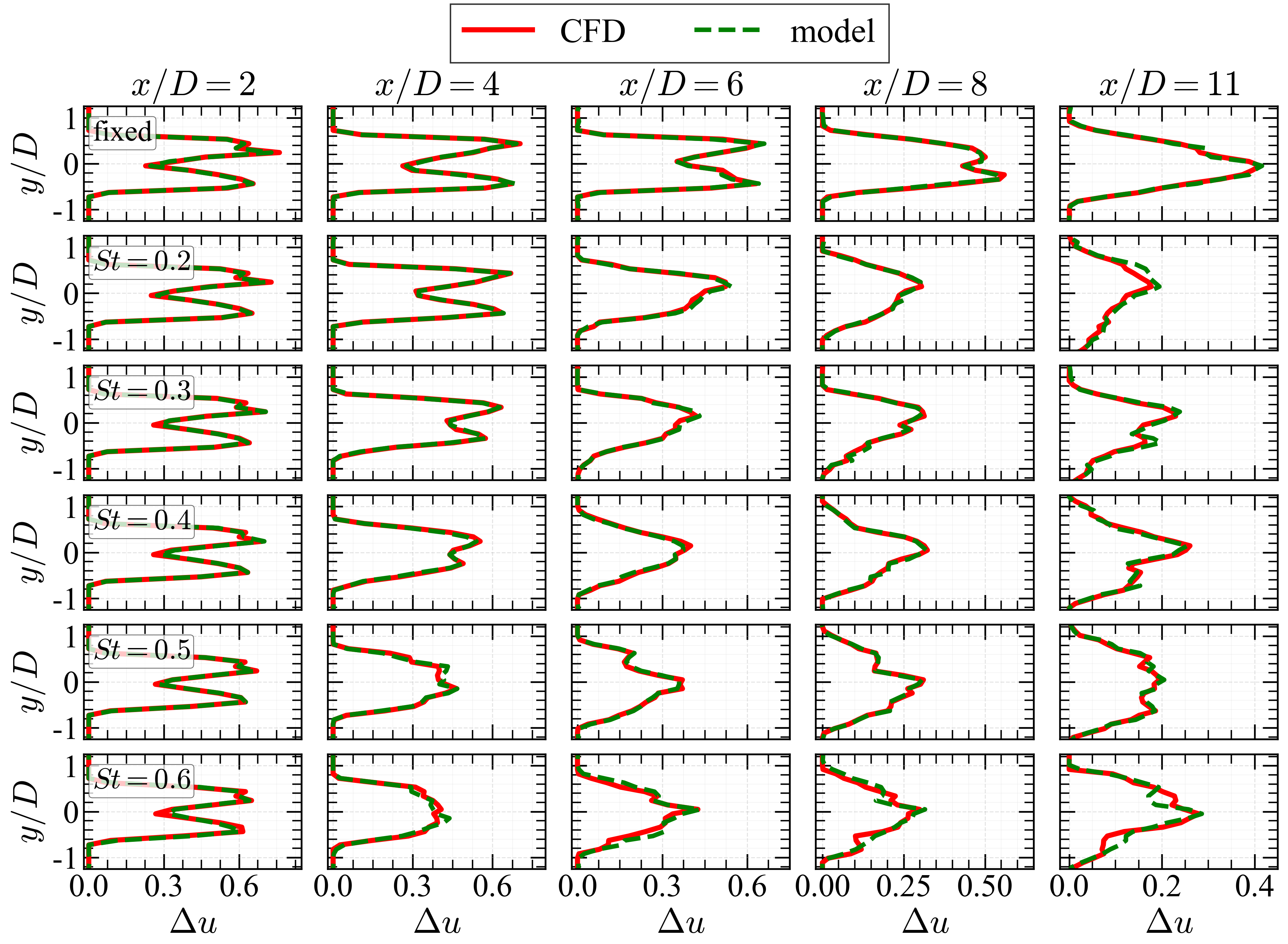}
\caption{Time-averaged spanwise wake-deficit profiles on the hub-height plane for the NREL 5~MW turbine.}
\label{fig:nrel_mean_spanwise_profiles}
\end{figure}

\begin{figure}[pos=!htbp,align=\centering,abovecap=0pt]
\centering
\includegraphics[width=0.82\textwidth, keepaspectratio]{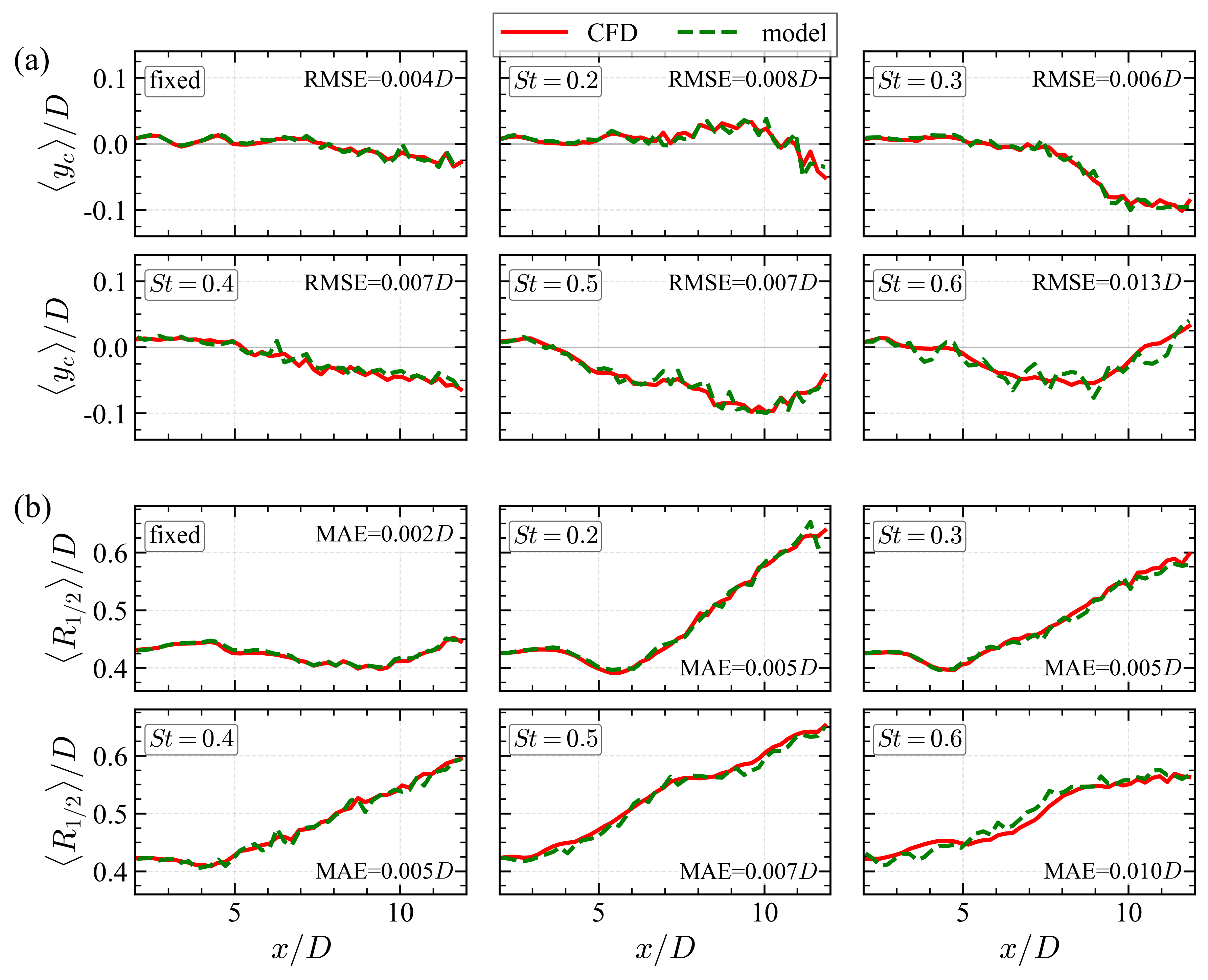}
\caption{Time-averaged wake evolution of the NREL 5~MW turbine for the fixed case and $St=0.2$--$0.6$: (a) wake center $\langle y_c\rangle/D$ with spatial RMSE values and (b) half-width $\langle R_{1/2}\rangle/D$ with spatial MAE values.}
\label{fig:nrel_mean_wake_geometry}
\end{figure}

When extrapolated beyond the training frequency envelope to the $St=0.6$ validation case, hub-height, center-plane, and 3D volumetric errors rise only modestly to 0.0419, 0.0369, and 0.0684, respectively. Contour comparison in Fig.~\ref{fig:v6slice} confirms that PIFNO preserves both the wake core and the convective motion-induced velocity modulation without excessive phase drift. Time-averaged spanwise wake-deficit profiles at five representative stations ($x/D=2,4,6,8,11$, Fig.~\ref{fig:nrel_mean_spanwise_profiles}) demonstrate that the operator faithfully captures the transition from the near-wake double-peak deficit induced by the root and hub to a smoother single-peak far-wake profile, exhibiting only mild deviations along the wake boundaries and centerline around $x/D=6$--$8$ at $St=0.6$. Along the streamwise distribution ($2\le x/D\le 12$), time-averaged wake-center trajectories (Fig.~\ref{fig:nrel_mean_wake_geometry}(a)) match CFD closely with spatial RMSEs of $0.004D$--$0.013D$, while wake half-widths (Fig.~\ref{fig:nrel_mean_wake_geometry}(b)) reproduce both the gradual spreading of the fixed rotor and the intensified motion-driven wake expansion with MAEs ranging from $0.002D$ (fixed) to $0.010D$ ($St=0.6$).

\begin{figure}[pos=!htbp,align=\centering,abovecap=0pt]
\centering
\includegraphics[width=0.82\textwidth,keepaspectratio]{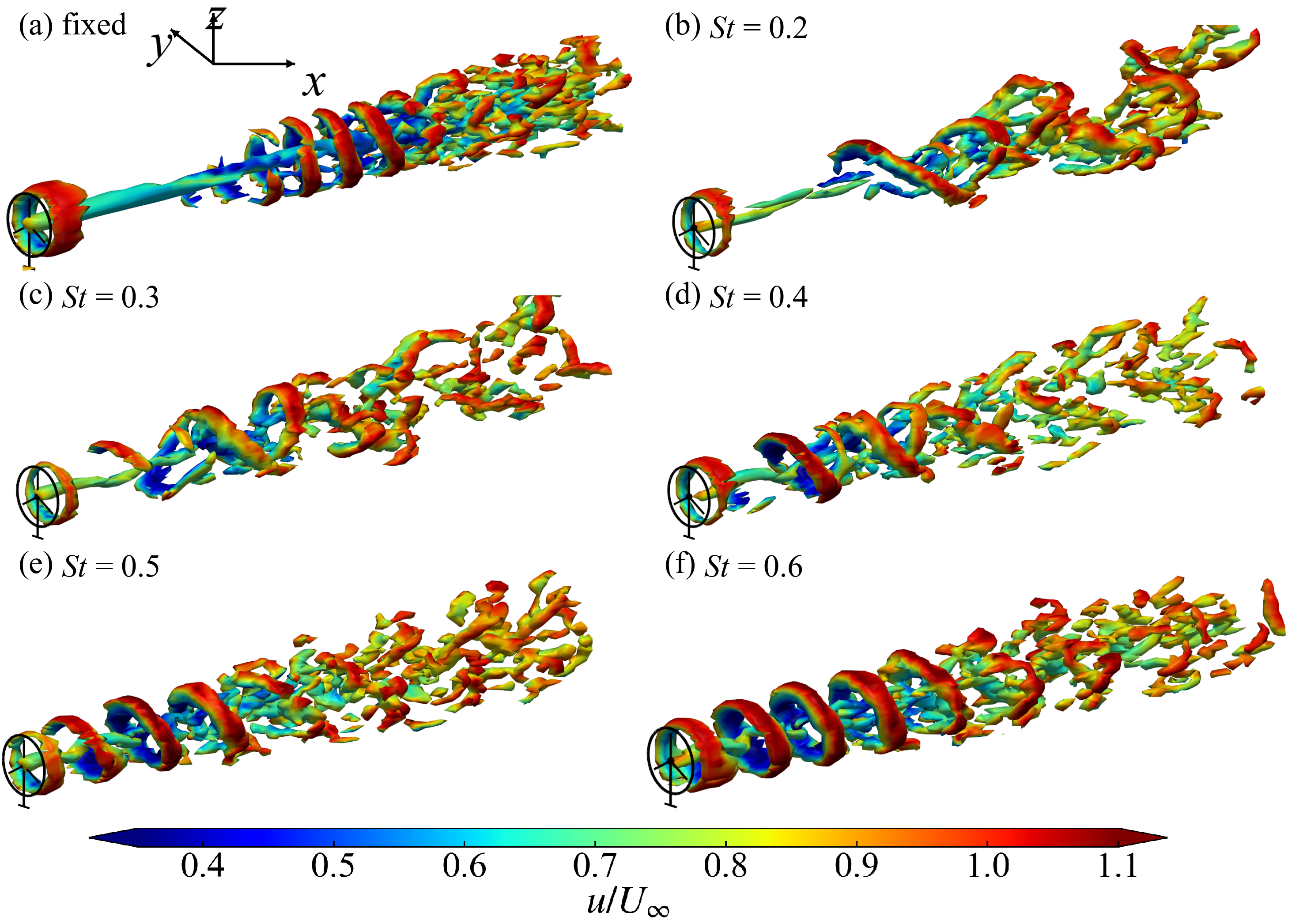}
\caption{Instantaneous 3D $Q$-criterion isosurfaces from CFD for the NREL 5~MW turbine across (a) fixed and (b--f) $St=0.2$--$0.6$ cases, colored by $u/U_\infty$.}
\label{fig:q_nrel_cfd}
\end{figure}

\begin{figure}[pos=!htbp,align=\centering,abovecap=0pt]
\centering
\includegraphics[width=0.82\textwidth,keepaspectratio]{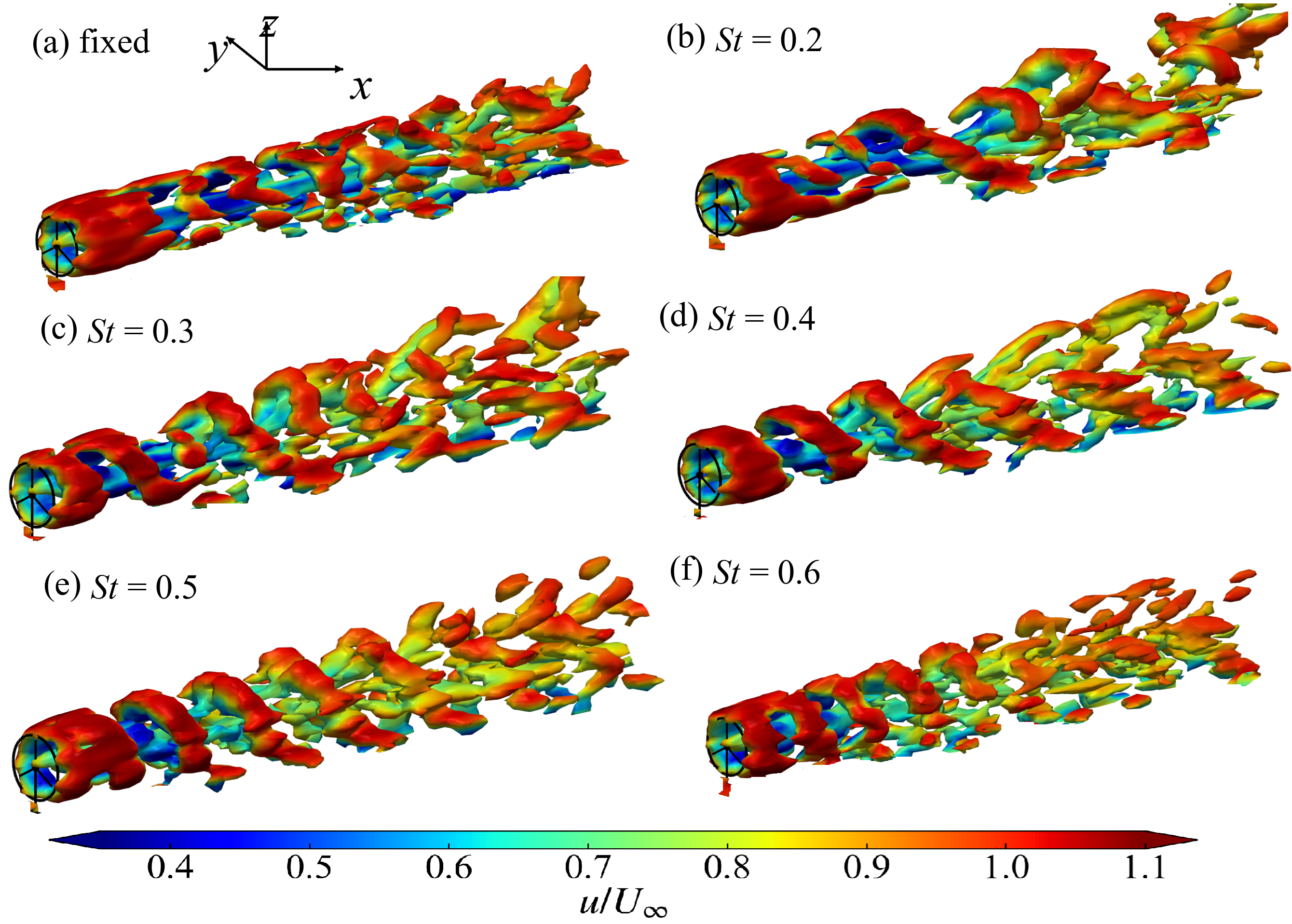}
\caption{Instantaneous 3D $Q$-criterion isosurfaces reconstructed by PIFNO for the NREL 5~MW turbine across (a) fixed and (b--f) $St=0.2$--$0.6$ cases, colored by $u/U_\infty$ using the same viewpoint and color scale as Fig.~\ref{fig:q_nrel_cfd}.}
\label{fig:q_nrel_pino}
\end{figure}

A comparison of instantaneous 3D $Q$-criterion isosurfaces between the LES--ALM reference (Fig.~\ref{fig:q_nrel_cfd}) and the PIFNO reconstructions (Fig.~\ref{fig:q_nrel_pino}) clarifies how well vortical coherence is retained off the observed planes. Across both the fixed baseline and all surge--pitch frequencies ($St=0.2$--$0.6$), PIFNO recovers the primary near-wake vortex rings, the cyclic vortex bunching induced by turbine motion, and the global helical wake envelope, while fine-scale fragmented vortices in the breaking far shear layer are noticeably smoothed. Because $Q$ depends quadratically on first-order spatial velocity derivatives, accurate integrated $L_2$ norms and mean wake-profile moments do not automatically guarantee sharp small-scale vorticity reconstruction, reflecting the combined influence of incomplete off-plane 3D measurements, finite Fourier mode truncation, and $L_2$-dominated supervision.

\subsection{Wake reconstruction from sparse planar observations}
\label{sec:sparse_wake_reconstruction}
To test planar observability when continuous spatial measurements are unavailable, fixed-weight sparse inference is evaluated across random coverages of 2\%, 5\%, 10\%, 25\%, and 50\% on the $St=0.6$ case under complete training supervision and idealized subsampled planar data (corresponding hub-height velocity fields appear in Figs.~\ref{fig:sparseratios} and~\ref{fig:iea_sparse_ratios}, \ref{app:additional_wake_diagnostics}). Across 2\%--50\% coverage, 3D streamwise-velocity errors reach 0.0768, 0.0733, 0.0689, 0.0705, and 0.0900 for the NREL 5~MW turbine, and 0.0767, 0.0736, 0.0716, 0.0808, and 0.1119 for the fully transferred IEA 22~MW model, respectively. Notably, 10\% NREL coverage nearly matches dense planar reconstruction (0.0689 compared with 0.0684 for dense sampling), and even 2\%--5\% sampling keeps volumetric error within roughly 0.077.

\begin{figure}[pos=!htbp,align=\centering,abovecap=0pt]
\centering
\includegraphics[width=0.75\textwidth,keepaspectratio]{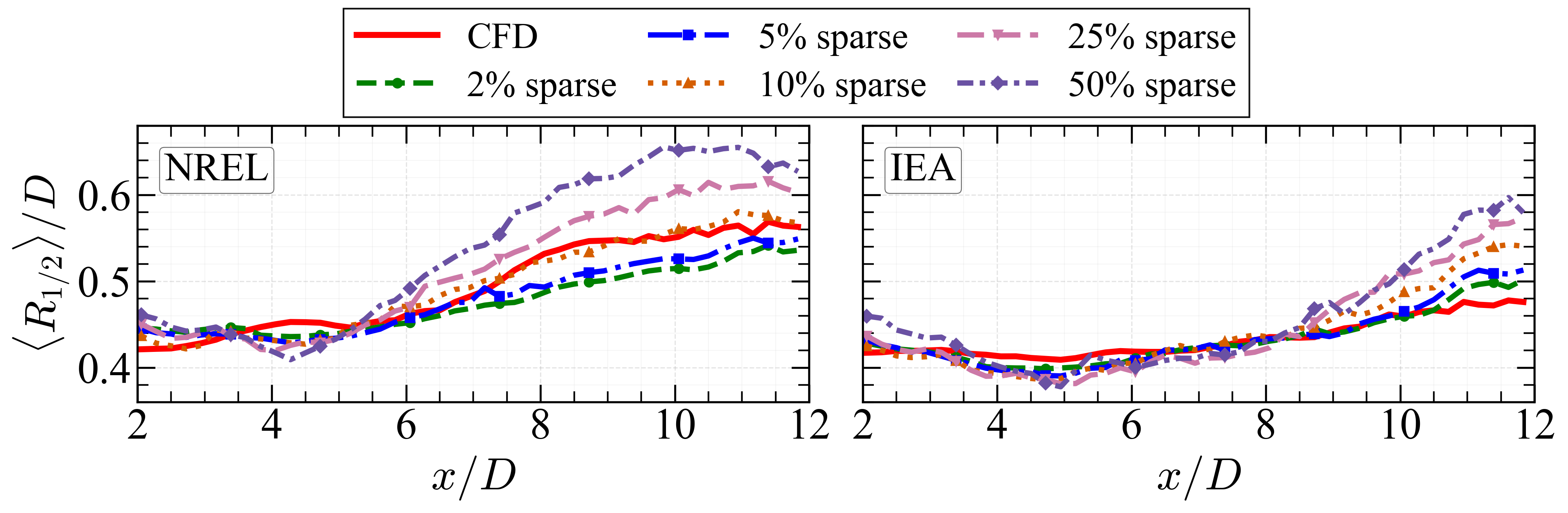}
\caption{Downstream evolution of the time-averaged wake half-width $\langle R_{1/2}\rangle/D$ under sparse planar observations at $St=0.6$ for the NREL 5~MW (left) and IEA 22~MW (right) turbines, comparing the CFD reference (red solid) against random coverages of 2\%, 5\%, 10\%, 25\%, and 50\%.}
\label{fig:sparse_half_width}
\end{figure}

Rather than decreasing monotonically with planar density, volumetric error attains its minimum at 10\% before rising at 25\% and 50\%. This non-monotonic behavior arises because sparse training draws mask coverages log-uniformly between 2\% and 50\% with checkpoint selection evaluated at the 10\% midpoint, anchoring the mask-conditioned operator response around 10\% sparsity. When frozen weights are subsequently queried at denser coverages (25\% and 50\%), uncalibrated mask conditioning shifts predictions away from the selected operating point. Consistently, Fig.~\ref{fig:sparse_half_width} shows that 10\% coverage tracks CFD wake-half-width evolution accurately on NREL 5~MW, whereas 25\% and 50\% inputs progressively overpredict downstream wake spreading for both turbine scales.

\subsection{Zero-shot cross-turbine wake reconstruction}

\begin{figure}[pos=!htbp,align=\centering,abovecap=0pt]
\centering
\includegraphics[width=0.76\textwidth,keepaspectratio]{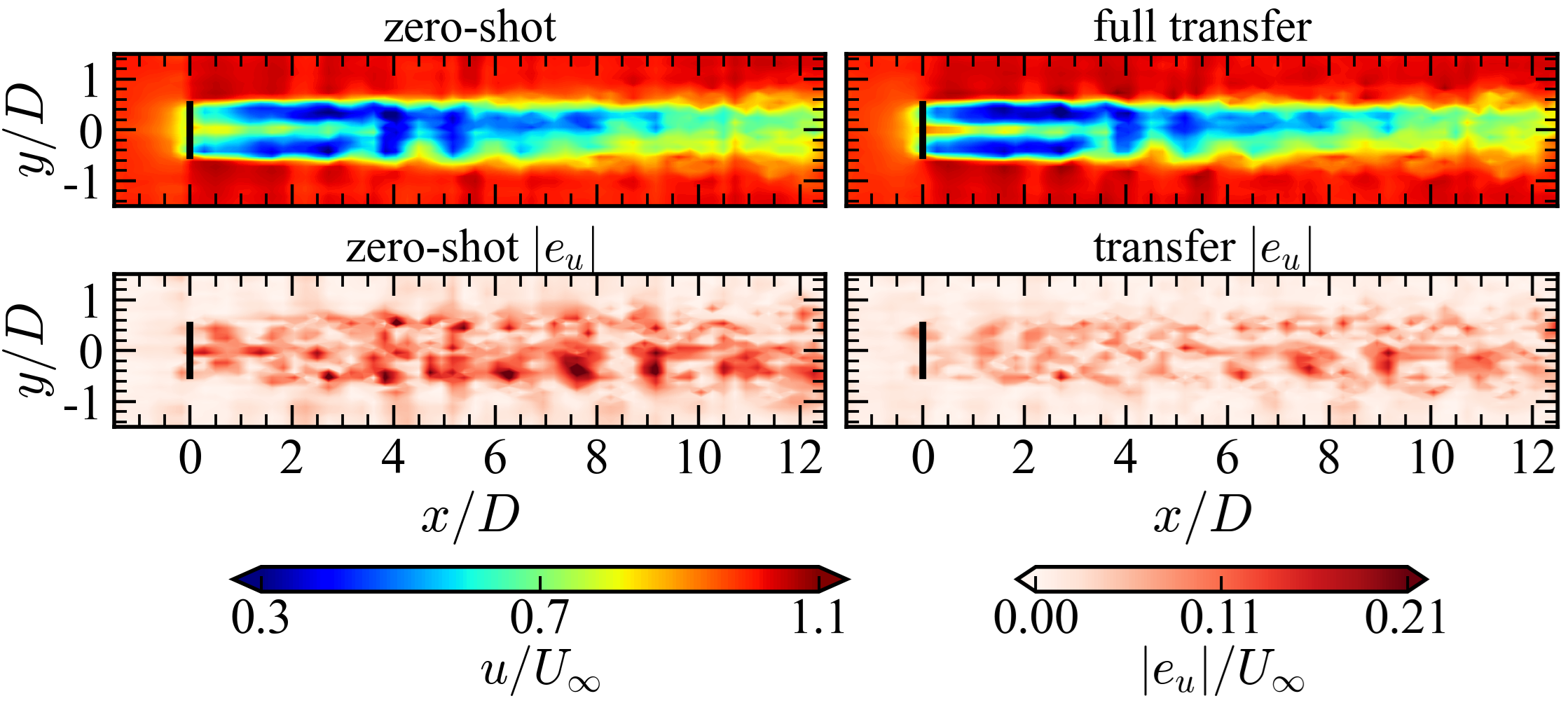}
\caption{Hub-height streamwise velocity $u/U_\infty$ (top) and absolute normalized error $|e_u|/U_\infty$ (bottom) for the IEA 22~MW turbine at $St=0.6$.}
\label{fig:transferslice}
\end{figure}

\begin{figure}[pos=!htbp,align=\centering,abovecap=0pt]
\centering
\includegraphics[width=0.74\textwidth,keepaspectratio]{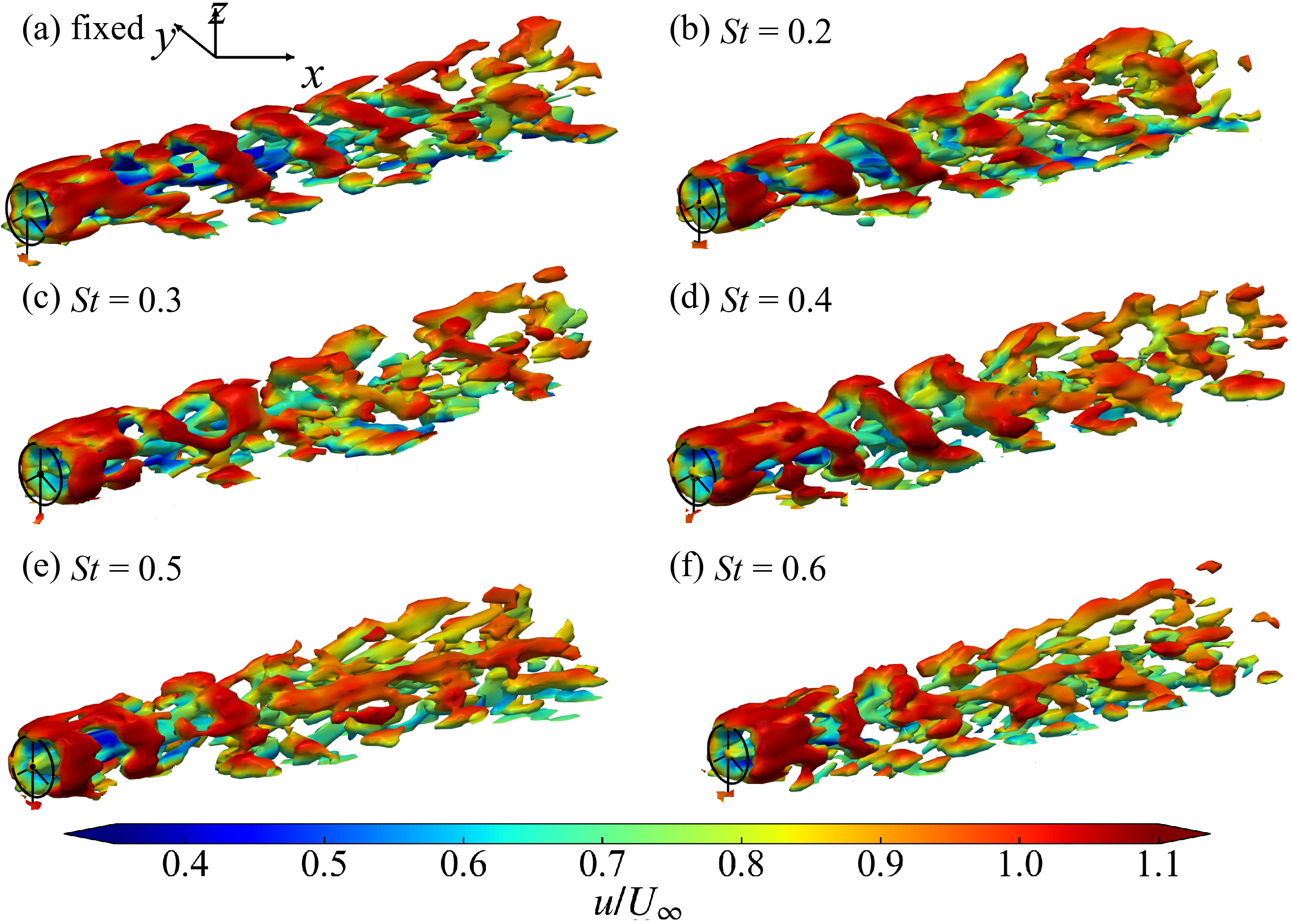}
\caption{Instantaneous 3D $Q$-criterion isosurfaces for the IEA 22~MW turbine reconstructed via zero-shot inference.}
\label{fig:q_iea_zeroshot}
\end{figure}

Applying the frozen NREL 5~MW operator directly to IEA 22~MW planar observations assesses zero-shot cross-scale generalization when rotor diameter (from 126 to 284~m), hub height (from 90 to 170~m), blade aerodynamic twist/chord schedule, and operating tip-speed ratio (from 8.0 to 9.1) all change concurrently. Under IEA $St=0.6$ forcing, relative errors climb to 0.0623 on the hub-height plane, 0.0475 on the centerline plane, and 0.1066 across the 3D streamwise velocity field (compared against 0.0684 for NREL 5~MW). Although nondimensionalization enables the zero-shot model to preserve the broad downstream deficit envelope and macroscopic wave phase in Fig.~\ref{fig:transferslice} (left), pronounced errors emerge along both shear layers and inside the near-wake core. Likewise, zero-shot 3D $Q$-criterion isosurfaces (Fig.~\ref{fig:q_iea_zeroshot}) capture the coarse outer envelope of wake undulation relative to the IEA CFD reference (Fig.~\ref{fig:q_iea_cfd}, \ref{app:additional_wake_diagnostics}), yet fail to resolve fine-scale vortex breakdown and rotor-specific tip-vortex spacing shifts associated with the higher TSR and lower $z_{\rm hub}/D$ ratio of the 22~MW rotor.

\subsection{Cross-turbine wake reconstruction with transfer learning}

\begin{figure}[pos=!htbp,align=\centering,abovecap=0pt]
\centering
\includegraphics[width=0.74\textwidth,keepaspectratio]{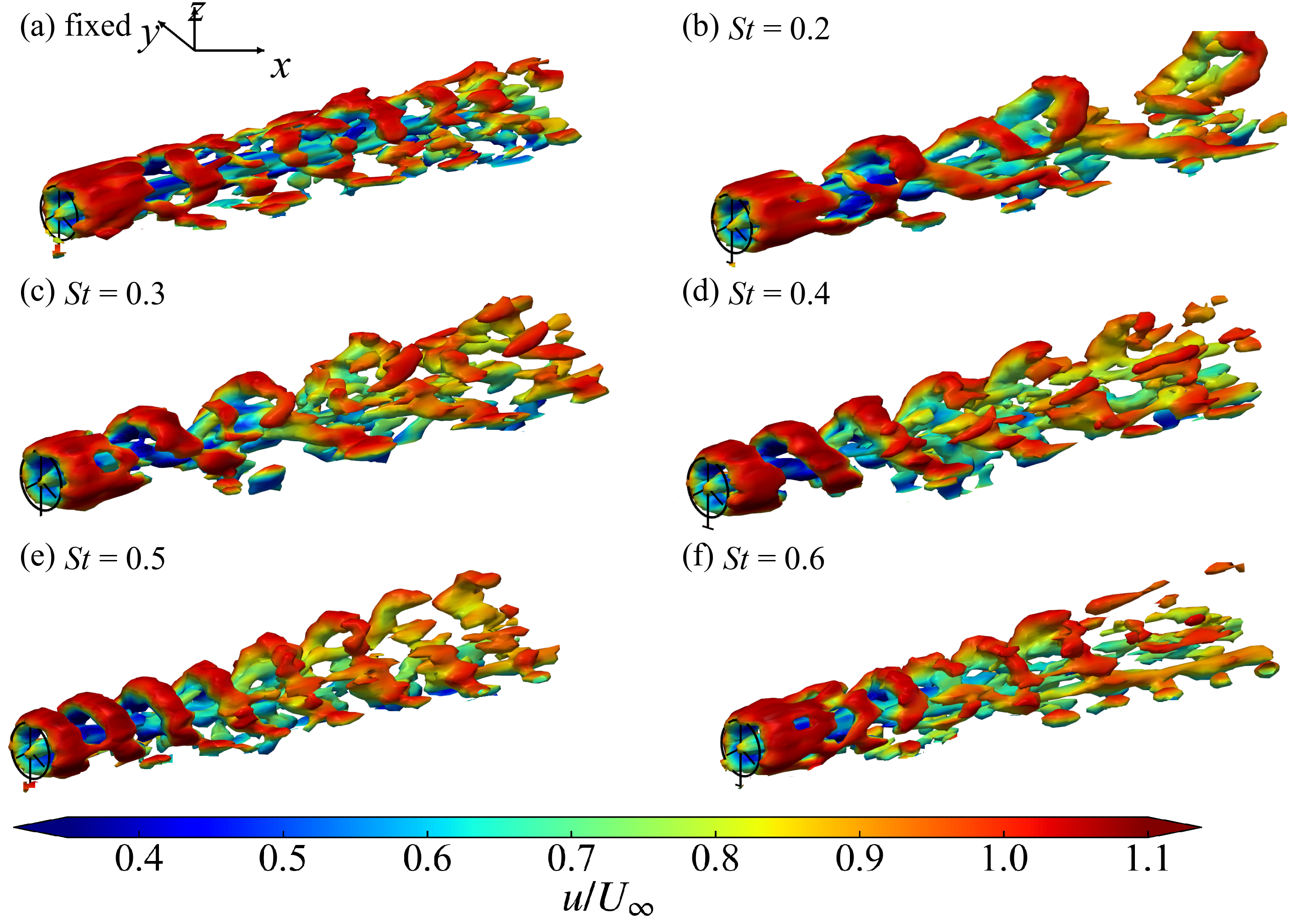}
\caption{Instantaneous 3D $Q$-criterion isosurfaces for the IEA 22~MW turbine reconstructed after full transfer learning.}
\label{fig:q_iea_transfer}
\end{figure}

Fine-tuning the source NREL operator on IEA 22~MW fixed and $St=0.2$--$0.5$ cases substantially improves cross-scale accuracy, lowering $St=0.6$ relative errors to 0.0383 on the hub-height plane, 0.0332 on the center plane, and 0.0682 across the 3D fluid volume, representing a 36\% reduction in 3D volumetric error relative to zero-shot scaling (0.1066) and restoring accuracy to the NREL baseline level (0.0684). As shown in Fig.~\ref{fig:transferslice} (right), transfer adaptation removes systematic deficit biases across the near-wake shear layers and core, while the reconstructed 3D $Q$-criterion structures (Fig.~\ref{fig:q_iea_transfer}) recover tighter, better-aligned near-wake vortex rings and coherent shear-layer shells than zero-shot inference, although fine-scale subgrid fragmentation remains attenuated relative to LES.

\begin{figure}[pos=!htbp,align=\centering,abovecap=0pt]
\centering
\includegraphics[width=0.7\textwidth,keepaspectratio]{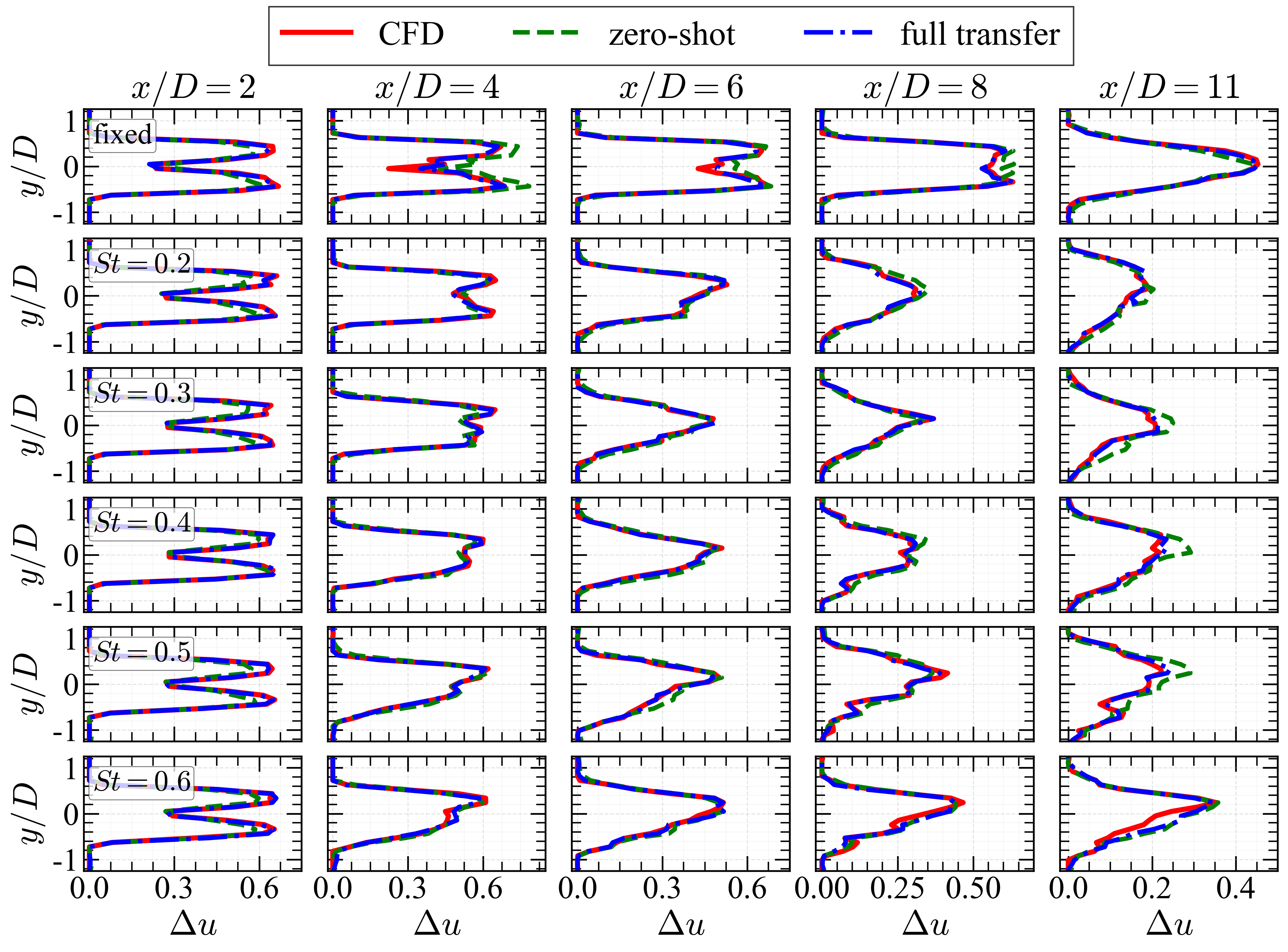}
\caption{Time-averaged spanwise wake-deficit profiles on the hub-height plane for the IEA 22~MW turbine.}
\label{fig:iea_mean_spanwise_profiles}
\end{figure}

\begin{figure}[pos=!htbp,align=\centering,abovecap=0pt]
\centering
\includegraphics[width=0.78\textwidth,keepaspectratio]{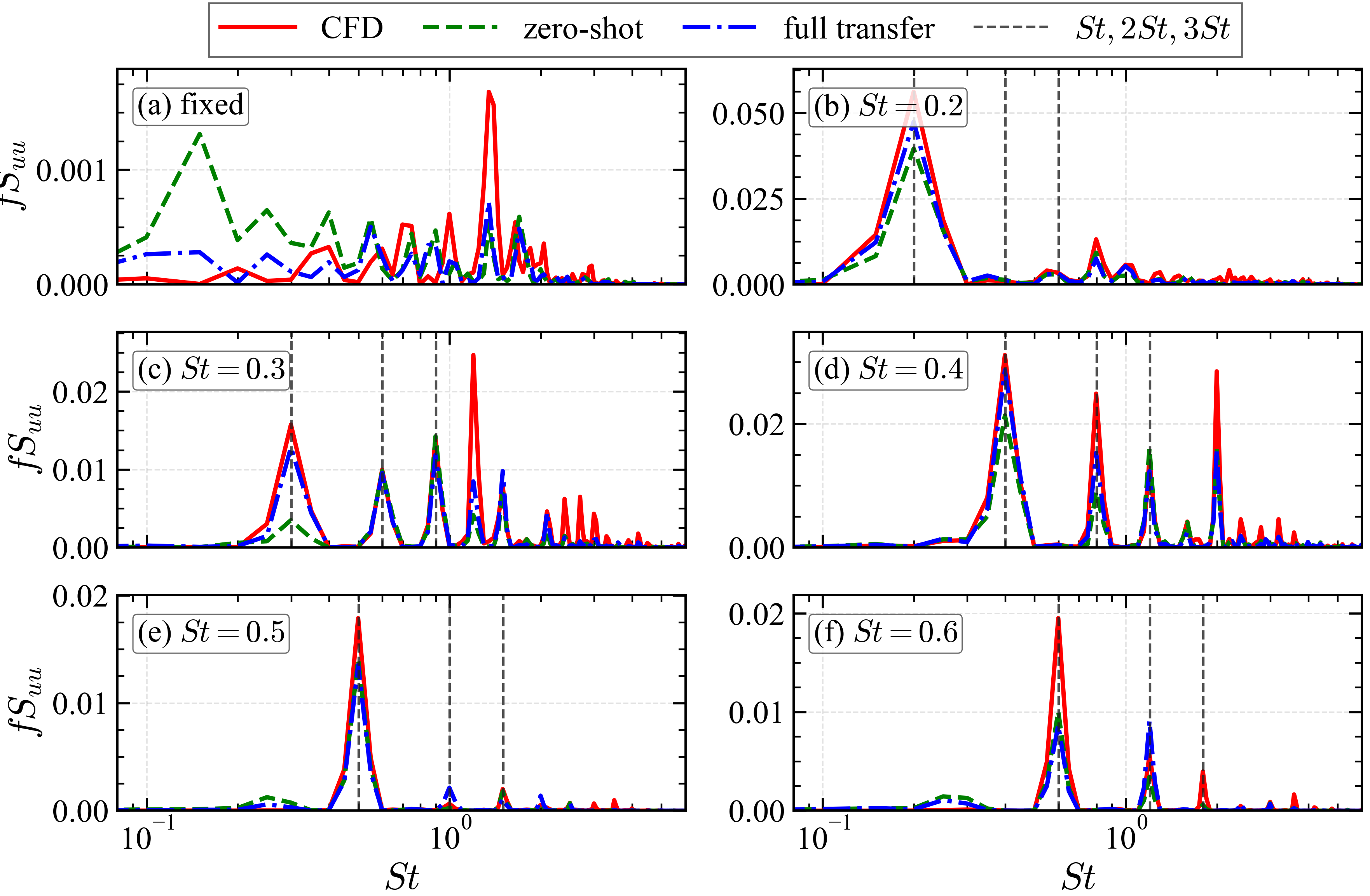}
\caption{Pre-multiplied power spectral density ($fS_{uu}$, equivalent to $\kappa S_{uu}(\kappa)$) of hub-height streamwise velocity fluctuations $u'$ near the wake centerline at $x/D=8$ for the IEA 22~MW turbine across (a) fixed and (b--f) $St=0.2$--$0.6$ cases.}
\label{fig:iea_fpsd_curves}
\end{figure}

Scale-resolved diagnostics clarify exactly which physical features transfer cleanly and which remain damped. In time-averaged spanwise deficit profiles across $x/D=2,4,6,8,11$ (Fig.~\ref{fig:iea_mean_spanwise_profiles}), full transfer substantially corrects centerline deficit depth and shoulder slopes relative to zero-shot predictions, most visibly in the fixed baseline and the far wake, while retaining modest localized deviations in intermediate wake stations. For spectral evaluation near $(x/D,y/D,z_*)=(8,0,0)$, the FPSD of streamwise velocity fluctuations $u'$ is computed using a full-record Hann window across the 500-sample observation window and plotted against the reduced frequency $\kappa=f_s D/U_\infty$, where $f_s S_{uu}(f_s)=\kappa S_{uu}(\kappa)$ (comparing model probe $(8.056,-0.048,-0.032)$ against CFD probe $(7.994,-0.010,0)$ in Fig.~\ref{fig:iea_fpsd_curves}). The spectra confirm that transfer adaptation recovers the dominant pre-multiplied fluctuation energy concentrated at the imposed forcing frequency $\kappa=St$ and its nonlinear harmonics ($2St$ and $3St$), with case-dependent differences in exact spectral peak amplitude and attenuated high-frequency broadband tails.

\begin{figure}[pos=!htbp,align=\centering,abovecap=0pt]
\centering
\includegraphics[width=0.8\textwidth,keepaspectratio]{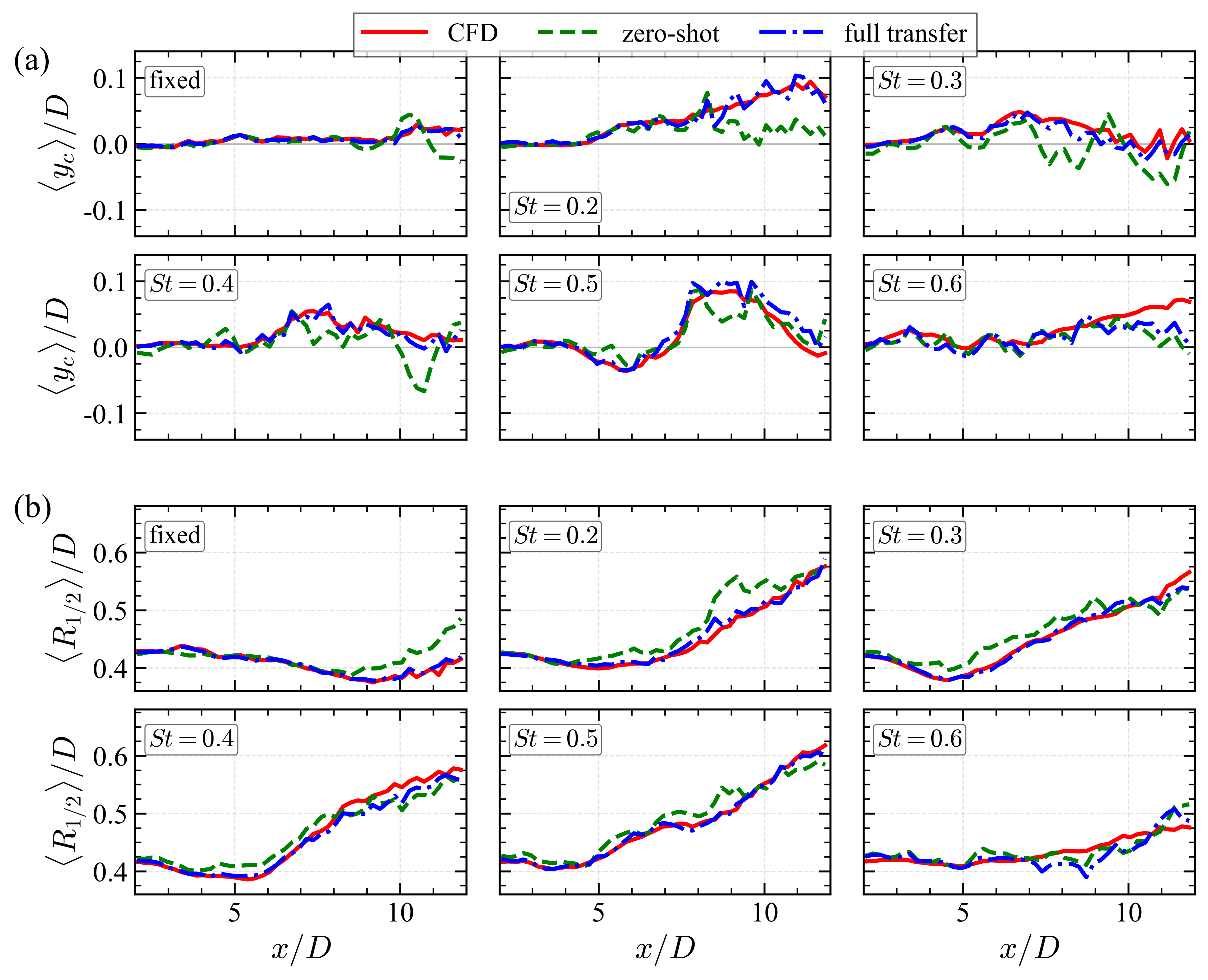}
\caption{Time-averaged wake evolution of the IEA 22~MW turbine: (a) wake center $\langle y_c\rangle/D$ and (b) half-width $\langle R_{1/2}\rangle/D$.}
\label{fig:iea_mean_wake_geometry}
\end{figure}

\begin{figure}[pos=!ht,align=\centering,abovecap=0pt]
\centering
\includegraphics[width=0.7\textwidth,keepaspectratio]{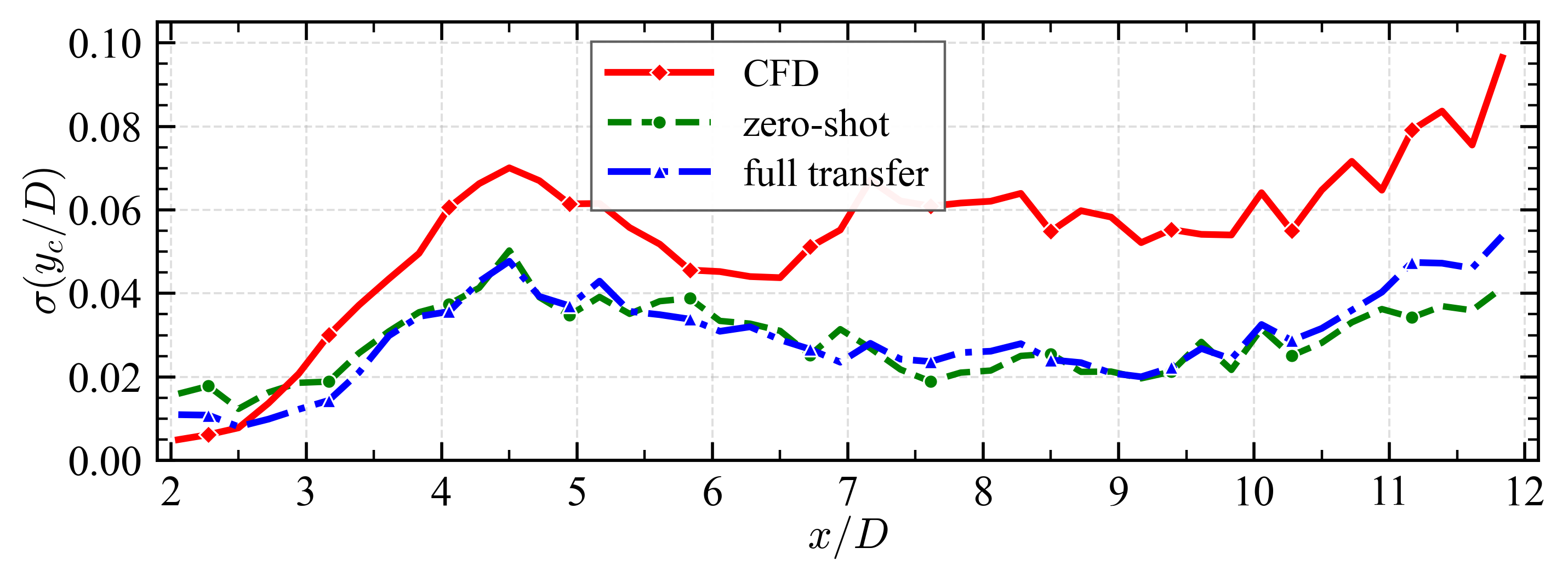}
\caption{Downstream evolution of the temporal standard deviation $\sigma(y_c/D)$ of the wake center for the IEA 22~MW turbine at $St=0.6$.}
\label{fig:iea_wake_center_std}
\end{figure}

Time-averaged wake-profile metrics, dynamic meandering diagnostics, and data-efficiency benchmarks reinforce this scale separation. Across all six regimes (Fig.~\ref{fig:iea_mean_wake_geometry}), full transfer consistently lowers time-averaged wake-center RMSE (from $0.033D$ to $0.009D$ at $St=0.2$ and from $0.025D$ to $0.020D$ at $St=0.6$) and improves moment-based half-width MAEs from fixed through $St=0.5$ (from $0.020D$ to $0.006D$ at $St=0.2$ and from $0.015D$ to $0.004D$ at $St=0.5$), whereas the moment-based mean half-width MAE ticks slightly upward at $St=0.6$ (from $0.011D$ to $0.013D$) because Gaussian fitting and deficit-moment metrics weight instantaneous profiles and time averaging differently. For unsteady wake dynamics on the IEA $St=0.6$ case, full transfer reduces the streamwise-averaged temporal RMSE of instantaneous Gaussian-fitted wake centers from $0.0481D$ to $0.0417D$ and decreases empirical half-width error from $0.0652D$ to $0.0495D$ (Fig.~\ref{fig:iea_wake_center_instantaneous}, \ref{app:additional_wake_diagnostics}). Nevertheless, the lateral meandering standard deviation $\sigma(y_c/D)$ remains underestimated below CFD (Fig.~\ref{fig:iea_wake_center_std}), reflecting partial damping of unforced turbulent wandering. Finally, the few-shot transfer benchmark (\ref{app:fewshot_transfer}) demonstrates that pre-training on NREL 5~MW cuts volumetric error by half compared with training from scratch when only one IEA case is available (0.0835 compared with 0.1693), while converging to the same asymptotic accuracy once five target cases are supplied (0.0682 compared with 0.0683).

\clearpage

\Needspace{8\baselineskip}
\section{Conclusions}
\label{sec:conclusions}

In this study, a physics-informed Fourier neural operator (PIFNO) framework is developed to reconstruct multiscale three-dimensional unsteady wakes of floating offshore wind turbines directly from two orthogonal observation planes by integrating space--time spectral operator learning with incompressible continuity and rotor-localized momentum constraints. Evaluated on high-fidelity wake fields generated via large-eddy simulations coupled with an actuator line model (LES--ALM) for the NREL 5~MW (rotor diameter $D = 126$~m) and IEA 22~MW ($D = 284$~m) turbines across fixed and surge--pitch cases spanning Strouhal numbers $St = fD/U_\infty = 0.2$--$0.6$ (where $f$ is turbine oscillation frequency, $U_\infty$ is inflow velocity, surge amplitude $A_s = 0.04D$, and pitch amplitude $A_p = 5^\circ$), the PIFNO model trained solely on the small-scale NREL 5~MW fixed and $St = 0.2$--$0.5$ regimes generalizes accurately to out-of-training forcing at $St = 0.6$. It achieves a 0.0684 volumetric streamwise-velocity relative $L_2$ error while recovering near-wake double-peak deficits, wake expansion, and coherent near-wake vortices, with mean wake-center root-mean-square errors within $0.013D$ and half-width mean absolute errors within $0.010D$. Under sparse $St = 0.6$ sampling, retaining only 10\% of planar points preserves near-dense three-dimensional accuracy for both the 5~MW (0.0689 compared with 0.0684 for dense sampling) and transferred 22~MW (0.0716 compared with 0.0682) turbines.

Building on the pre-trained source operator, cross-scale transfer learning bridges small-scale 5~MW and large-scale 22~MW rotor wakes while substantially reducing target simulation demands relative to zero-shot scaling and training from scratch. At 22~MW $St = 0.6$, transfer learning cuts volumetric error by 36\% versus zero-shot inference (0.1066 to 0.0682), halves one-shot scratch error with a single fixed target case (0.1693 to 0.0835), and lowers three-shot error from 0.1038 to 0.0745. Coordinated diagnostics spanning time-averaged profiles, instantaneous wake-center trajectories ($0.0481D$ to $0.0417D$), empirical half-width error ($0.0652D$ to $0.0495D$), pre-multiplied velocity spectra (restoring harmonics near $St$, $2St$, and $3St$), and three-dimensional $Q$-criterion vortices distinguish transferable large-scale motion-driven wake dynamics from damped fine-scale breakdown. Overall, the proposed physics-informed operator and cross-scale transfer framework demonstrates strong potential for rapid three-dimensional wake reconstruction, downstream inflow assessment, and wake-control optimization in utility-scale floating offshore wind farms.

\section*{Acknowledgements}
This work was supported by the National Natural Science Foundation of China (nos. 52606283, 12202456, 52406234), the Natural Science Foundation of Jiangsu Province (no. BK20250336), the China Postdoctoral Science Foundation (no. 2025M773143), and the Jiangsu Young Elite Scientist Sponsorship Project for Science and Technology (no. JSTJ-2026-517).

\section*{Declaration of competing interest}\label{sec:competing-interests}
The authors report no conflict of interest.

\section*{CRediT authorship contribution statement}
\textbf{Guodan Dong:} Conceptualization, Methodology, Software, Formal analysis, Investigation, Data curation, Visualization, Writing -- original draft.
\textbf{Qi Qian:} Data curation, Formal analysis.
\textbf{Feifei Xue:} Data curation, Funding acquisition, Writing -- review \& editing.
\textbf{Jianhua Qin:} Conceptualization, Methodology, Project administration, Writing -- review \& editing.
\textbf{Chang Xu:} Conceptualization, Supervision, Writing -- review \& editing.
\textbf{Wenzhong Shen:} Methodology, Supervision, Writing -- review \& editing.

\section*{Data availability}
Data will be made available on request.

\section*{Declaration of generative AI and AI-assisted technologies in the writing process}
The authors declare that generative AI and AI-assisted technologies were not used in the writing process of this manuscript.

\FloatBarrier
\appendix
\setlength{\textfloatsep}{6pt plus 2pt minus 2pt}
\setlength{\floatsep}{6pt plus 2pt minus 2pt}
\setlength{\intextsep}{6pt plus 2pt minus 2pt}
\renewcommand{\thesection}{\appendixname~\Alph{section}}
\renewcommand{\thefigure}{\Alph{section}.\arabic{figure}}
\renewcommand{\thetable}{\Alph{section}.\arabic{table}}
\counterwithin{equation}{section}
\renewcommand{\theequation}{\Alph{section}.\arabic{equation}}
\setcounter{figure}{0}
\setcounter{table}{0}
\renewcommand*{\theHfigure}{appendix.\Alph{section}.\arabic{figure}}
\renewcommand*{\theHtable}{appendix.\Alph{section}.\arabic{table}}

\section{NREL 5~MW training and network ablation}
\label{app:nrel_ablation}

\begin{figure}[pos=!htbp,align=\centering,abovecap=0pt]
\centering
\includegraphics[width=0.78\textwidth,keepaspectratio]{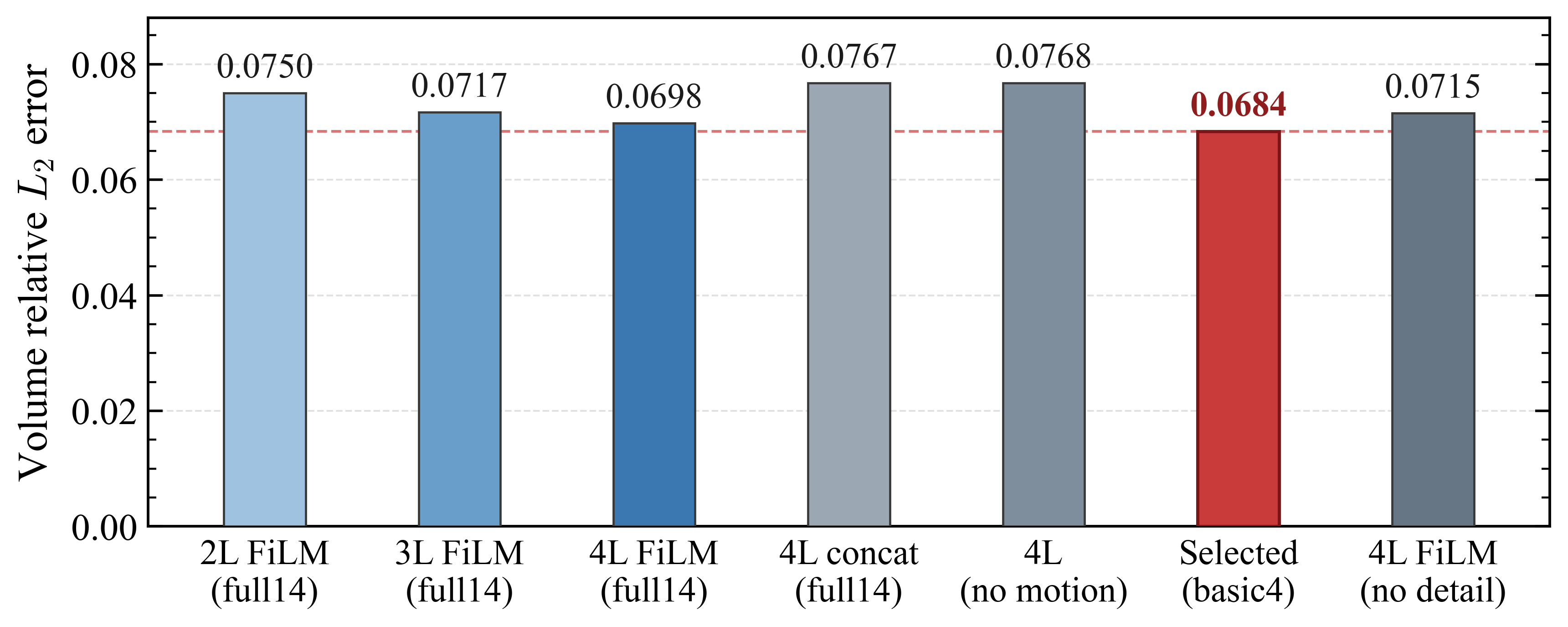}
\caption{Architectural ablation on the NREL 5~MW $St=0.6$ case corresponding to Table~\ref{tab:ablation}.}
\label{fig:ablation}
\end{figure}

Table~\ref{tab:ablation} and Fig.~\ref{fig:ablation} compare architecture variants on NREL $St=0.6$ under the shared residual objective. Whereas the selected model uses the compact \texttt{basic4} vector $\bm{s}_o$ (Eq.~\eqref{eq:motion_descriptor}), the 14-feature \texttt{full14} basis $\bm{\phi}_s$ in Eq.~\eqref{eq:residual_force} ($s_\psi=\sin\psi$, $c_\psi=\cos\psi$) is:
\begin{linenomath}
\begin{equation}
\begin{split}
\bm{\phi}_s=(&St,St^2,s_\psi,c_\psi,St\,c_\psi,-St\,s_\psi,
-St^2s_\psi,-St^2c_\psi,\\
&c_\psi,-s_\psi,-St\,s_\psi,-St\,c_\psi,-St^2c_\psi,St^2s_\psi)^T,
\end{split}
\label{eq:full14_basis}
\end{equation}
\end{linenomath}
where entries 9--14 duplicate phase-shifted combinations of entries 3--8, and \texttt{concat-FNO4D} supplies motion descriptors solely as input channels without FiLM.

\begin{table}[pos=htbp]
\centering
\caption{Architectural ablation on the NREL 5~MW validation case ($St=0.6$). Reported metrics are relative $L_2$ errors, where hub-height ($x$-$y$) planar errors include $u,v,w,p$ and 3D volume errors evaluate only $u$.}
\label{tab:ablation}
\small
\setlength{\tabcolsep}{0pt}
\renewcommand{\arraystretch}{1.12}
\begin{tabular*}{\textwidth}{@{\extracolsep{\fill}}lccccc@{}}
\toprule
Model variant & Layers & \shortstack{Motion\\input} & Detail & $x$-$y$ & \shortstack{3D\\volume} \\
\midrule
2-layer FiLM-FNO4D & 2 & full14 & yes & 0.0361 & 0.0750 \\
3-layer FiLM-FNO4D & 3 & full14 & yes & 0.0395 & 0.0717 \\
4-layer FiLM-FNO4D & 4 & full14 & yes & 0.0418 & 0.0698 \\
4-layer concat-FNO4D & 4 & full14 & yes & 0.0437 & 0.0767 \\
4-layer FiLM-FNO4D & 4 & none & yes & 0.0421 & 0.0768 \\
Selected FiLM-FNO4D & 4 & basic4 & yes & 0.0419 & 0.0684 \\
4-layer FiLM-FNO4D & 4 & full14 & no & 0.0540 & 0.0715 \\
\bottomrule
\end{tabular*}
\end{table}

The selected four-block FiLM operator with \texttt{basic4} and the residual detail decoder attains the lowest 3D error (0.0684), outperforming shallower blocks, direct channel concatenation (0.0767), unconditioned FNO4D (0.0768), and detail-decoder removal (0.0715).

\setcounter{figure}{0}
\setcounter{table}{0}

\section{Few-shot transfer learning and scratch baseline}
\label{app:fewshot_transfer}

\begin{figure}[pos=!htbp,align=\centering,abovecap=0pt]
\centering
\includegraphics[width=0.78\textwidth,keepaspectratio]{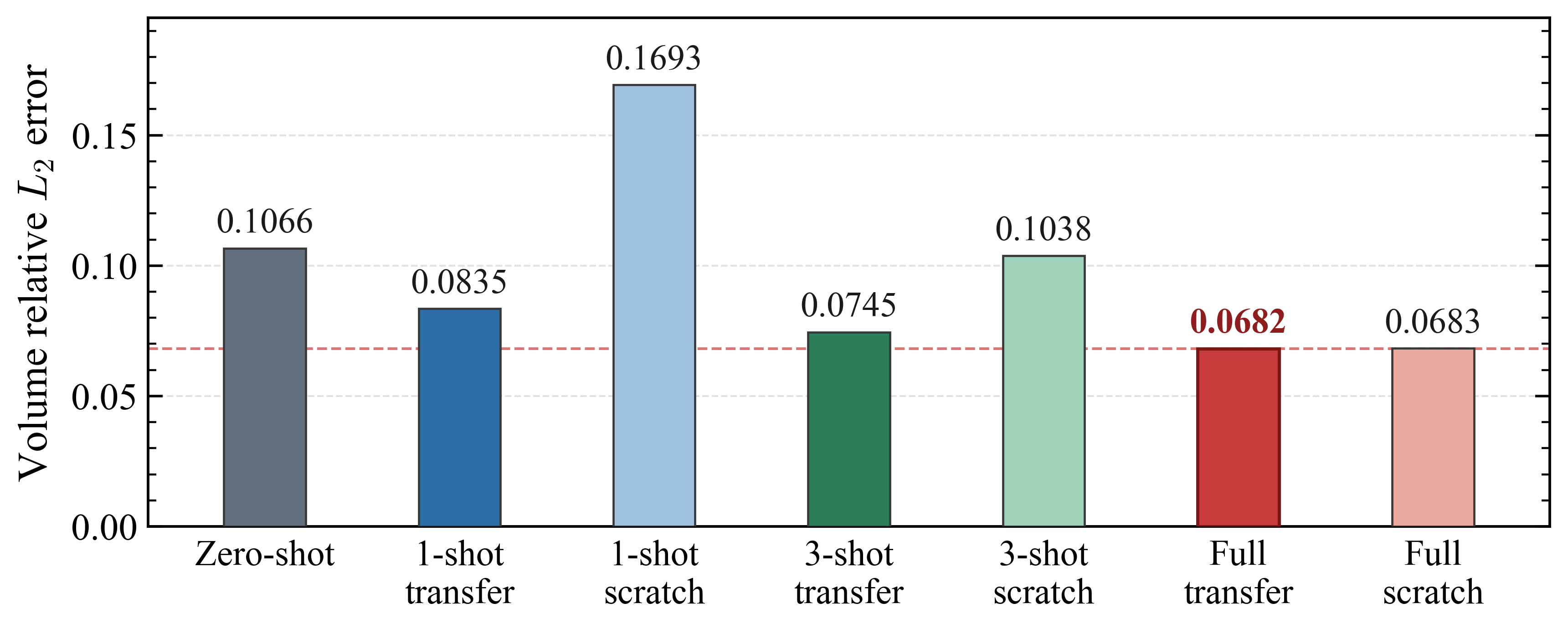}
\caption{Cross-turbine transfer learning versus scratch on the IEA 22~MW $St=0.6$ case corresponding to Table~\ref{tab:fewshot}.}
\label{fig:fewshot}
\end{figure}

Table~\ref{tab:fewshot} and Fig.~\ref{fig:fewshot} compare NREL-initialized transfer learning (60 epochs at $2\times 10^{-4}$) against scratch training (80 epochs at $5\times 10^{-4}$) on IEA $St=0.6$. Transfer markedly outperforms scratch under scarce target data (one-shot fixed: 0.0835 compared with 0.1693, three-shot: 0.0745 compared with 0.1038), while both reach 0.0682--0.0683 once all five cases are supplied.

\begin{table}[pos=htbp]
\centering
\caption{IEA 22~MW $St=0.6$ validation errors for transfer learning and random-initialization (scratch).}
\label{tab:fewshot}
\small
\setlength{\tabcolsep}{0pt}
\renewcommand{\arraystretch}{1.12}
\begin{tabular*}{\textwidth}{@{\extracolsep{\fill}}lcccc@{}}
\toprule
Target data and initialization & Epochs & \shortstack{Hub-height\\plane} & \shortstack{Center\\plane} & \shortstack{3D\\volume} \\
\midrule
zero-shot, NREL init & 0 & 0.0623 & 0.0475 & 0.1066 \\
fixed, NREL init & 60 & 0.0496 & 0.0425 & 0.0835 \\
fixed, scratch & 80 & 0.0915 & 0.0873 & 0.1693 \\
fixed, $St=0.2$, $St=0.3$, NREL init & 60 & 0.0418 & 0.0363 & 0.0745 \\
fixed, $St=0.2$, $St=0.3$, scratch & 80 & 0.0591 & 0.0599 & 0.1038 \\
fixed--$St=0.5$, NREL init & 60 & 0.0383 & 0.0332 & 0.0682 \\
fixed--$St=0.5$, scratch & 80 & 0.0412 & 0.0367 & 0.0683 \\
\bottomrule
\end{tabular*}
\end{table}

\setcounter{figure}{0}
\section{Additional wake reconstruction diagnostics}
\label{app:additional_wake_diagnostics}

\begin{figure}[pos=!htbp,align=\centering,abovecap=0pt]
\centering
\includegraphics[width=0.62\textwidth,keepaspectratio]{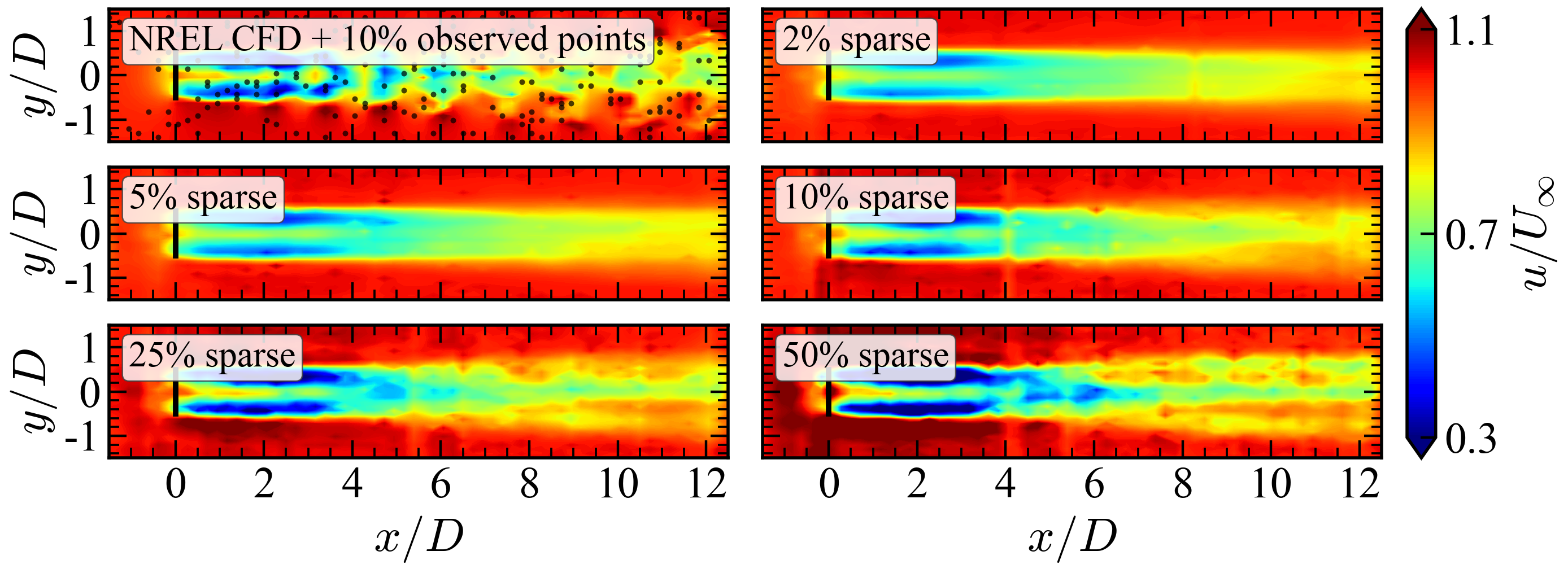}
\caption{Instantaneous hub-height $u/U_\infty$ fields for the NREL 5~MW turbine at $St=0.6$ under sparse observations.}
\label{fig:sparseratios}
\end{figure}

\begin{figure}[pos=!htbp,align=\centering,abovecap=0pt]
\centering
\includegraphics[width=0.62\textwidth,keepaspectratio]{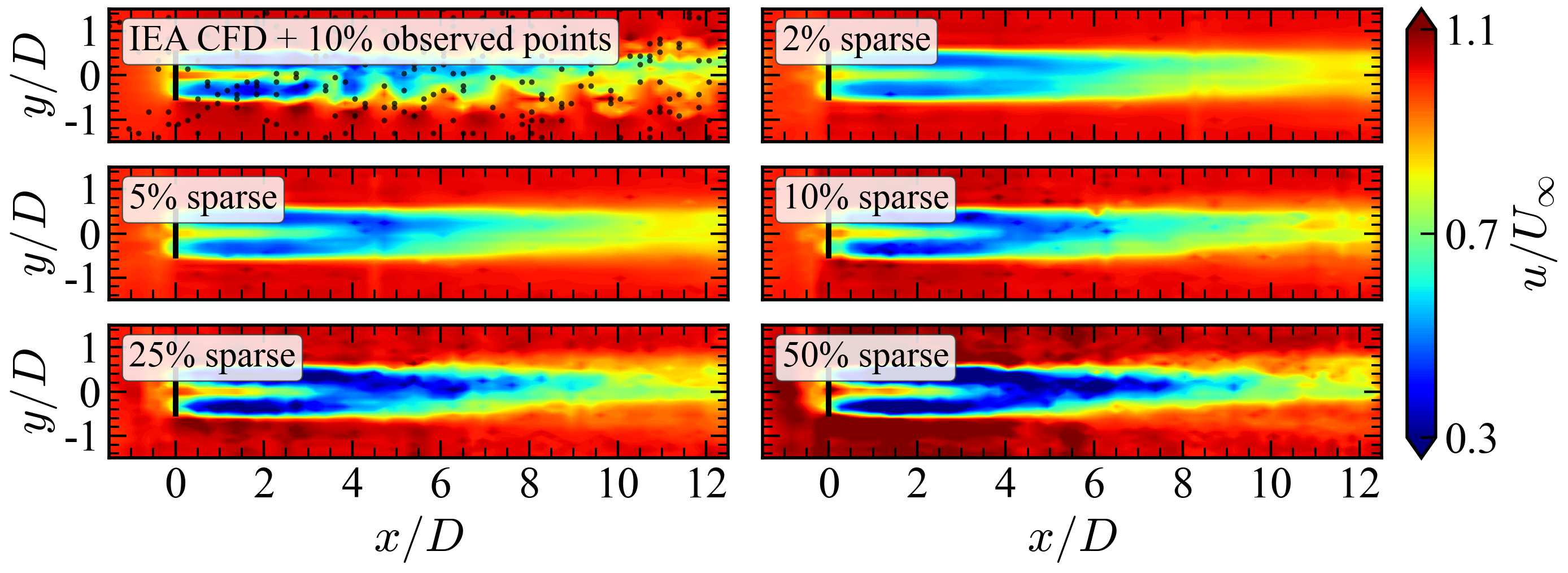}
\caption{Instantaneous hub-height $u/U_\infty$ fields for the IEA 22~MW turbine at $St=0.6$ under sparse observations.}
\label{fig:iea_sparse_ratios}
\end{figure}

\begin{figure}[pos=!htbp,align=\centering,abovecap=0pt]
\centering
\includegraphics[width=0.76\textwidth,keepaspectratio]{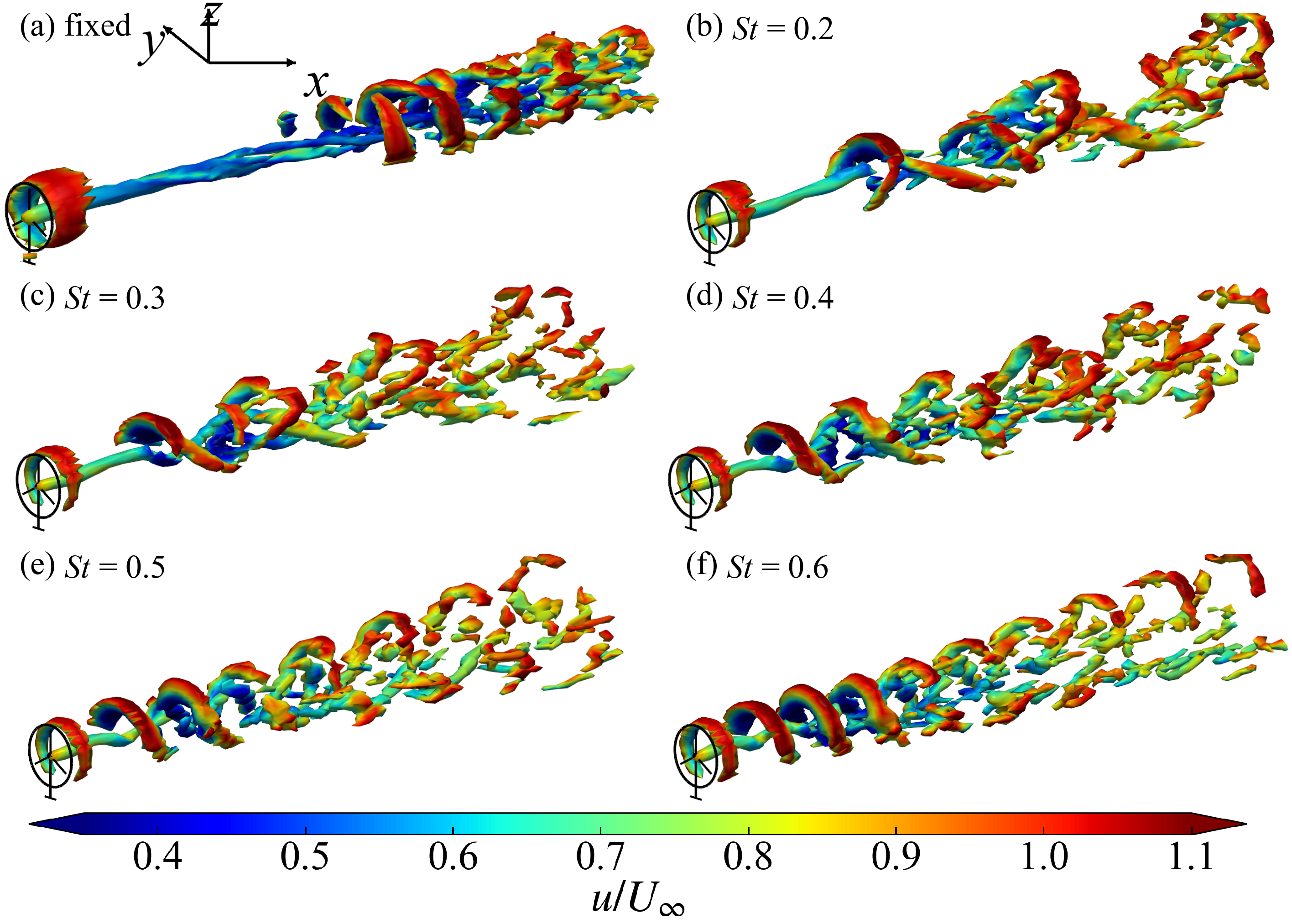}
\caption{Instantaneous 3D $Q$-criterion isosurfaces from CFD for the IEA 22~MW turbine, serving as the reference baseline for Figs.~\ref{fig:q_iea_zeroshot} and \ref{fig:q_iea_transfer}.}
\label{fig:q_iea_cfd}
\end{figure}

\begin{figure}[pos=!htbp,align=\centering,abovecap=0pt]
\centering
\includegraphics[width=0.84\textwidth,keepaspectratio]{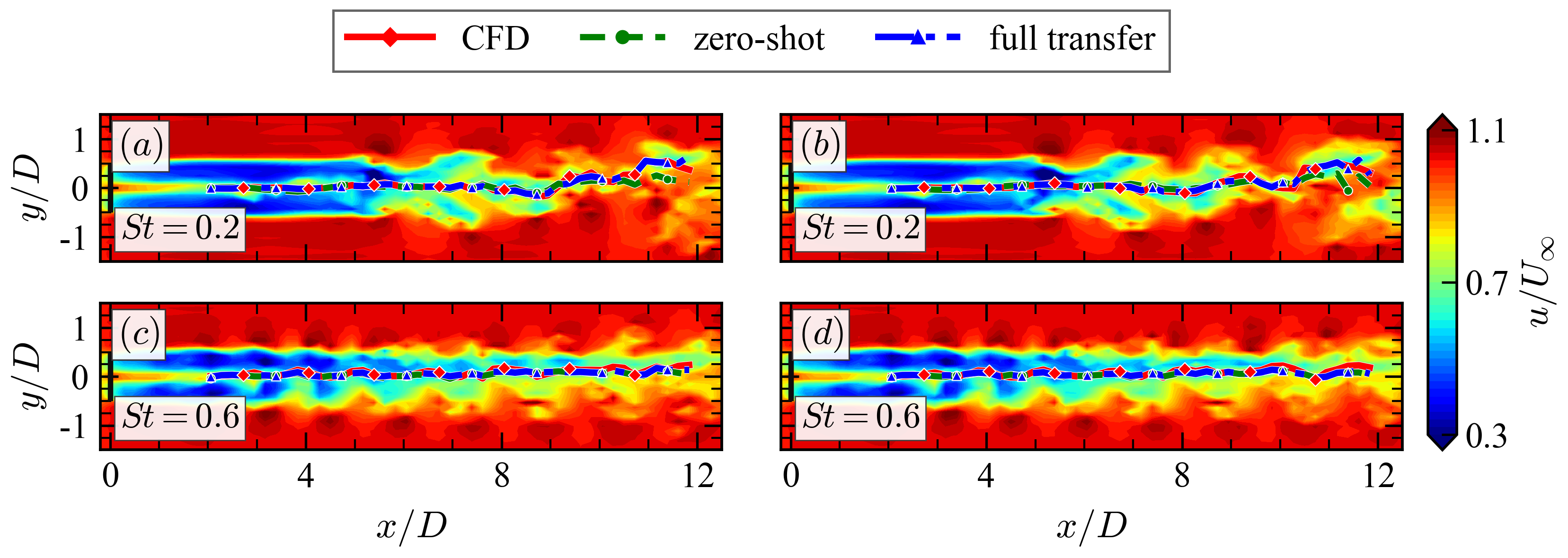}
\caption{Instantaneous Gaussian-fitted wake-center trajectories overlaid on IEA 22~MW CFD hub-height $u/U_\infty$ contours at two representative instants each for (a, b) $St=0.2$ and (c, d) $St=0.6$.}
\label{fig:iea_wake_center_instantaneous}
\end{figure}

Figures~\ref{fig:sparseratios} and \ref{fig:iea_sparse_ratios} show instantaneous hub-height $u/U_\infty$ fields across 2\%--50\% sparse coverage at $St=0.6$, where fixed-weight sparse inference at 25\% and 50\% produces an elongated low-speed IEA core while overpredicting wake width. Figures~\ref{fig:q_iea_cfd} and \ref{fig:iea_wake_center_instantaneous} give the IEA CFD $Q$-criterion baseline and instantaneous Gaussian-fitted wake-center trajectories ($2\le x/D\le 12$) for $St=0.2$ and $St=0.6$, confirming closer far-wake tracking after full transfer.

\FloatBarrier

\bibliographystyle{elsarticle-num}
\bibliography{references}

\end{document}

%% file: figure_annotations.tex
\definecolor{diagramblue}{rgb}{0.969,0.984,1}
\definecolor{diagramgreen}{rgb}{0.949,0.988,0.941}
\newcommand{\figannotationtext}[2]{\scalebox{#1}{\fontsize{10}{12}\selectfont #2}}
\newcommand{\workflowgraphic}{%
  \begingroup
  \setlength{\unitlength}{0.01\textwidth}%
  \setlength{\fboxsep}{0pt}%
  \begin{picture}(100,46.5632)
    \put(0,0){\includegraphics[width=\textwidth]{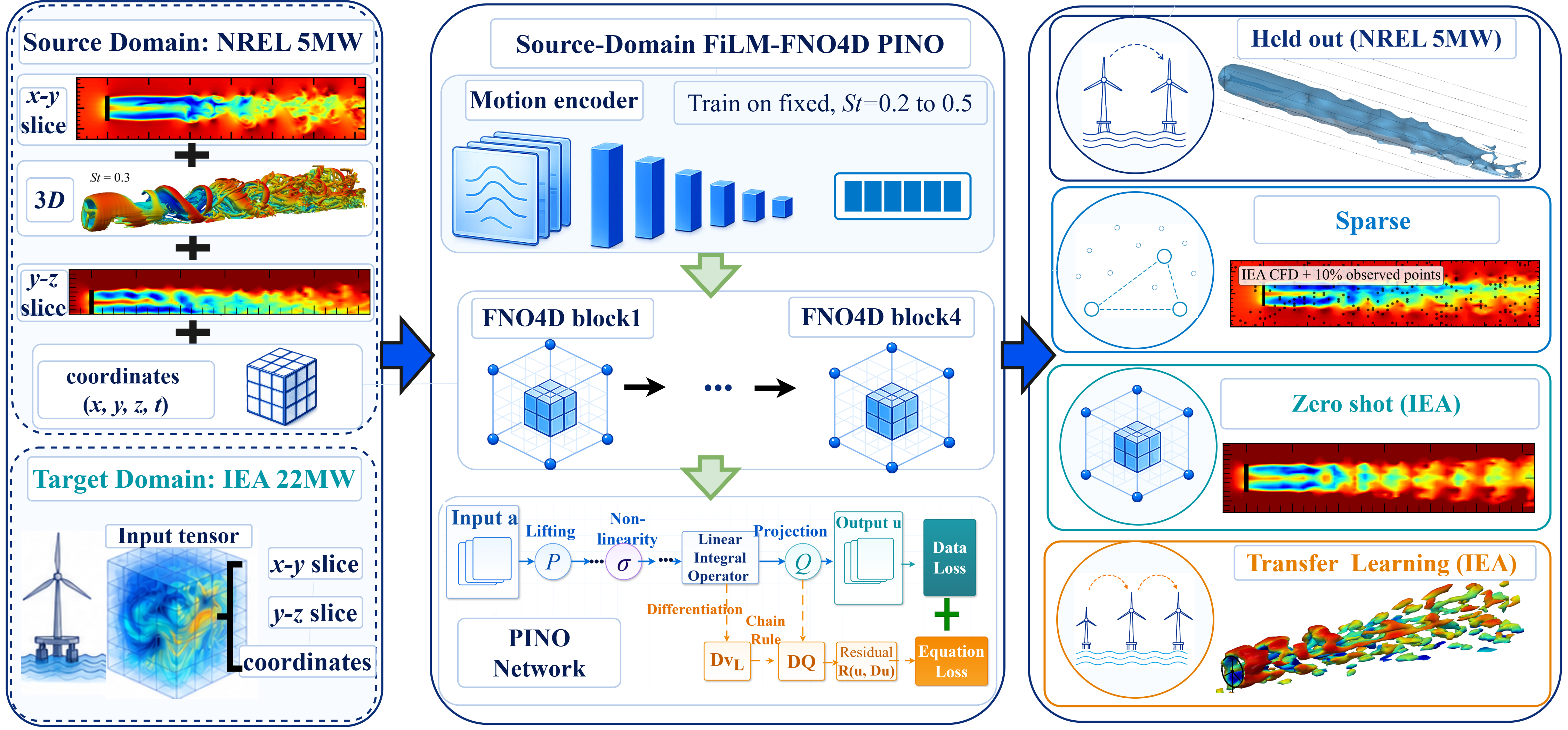}}
    \put(1.35,28.10){\colorbox{white}{\makebox(2.8,1.65){\figannotationtext{0.7}{$x$-$z$}}}}
    \put(17.05,6.85){\colorbox{white}{\makebox(2.8,1.65){\figannotationtext{0.7}{$x$-$z$}}}}
    \put(78.90,42.92){\colorbox{white}{\makebox(17.3,2.15){\figannotationtext{0.68}{Validation (NREL 5 MW)}}}}
  \end{picture}%
  \endgroup
}
\newcommand{\architecturegraphic}{%
  \begingroup
  \setlength{\unitlength}{0.01\textwidth}%
  \setlength{\fboxsep}{0pt}%
  \begin{picture}(100,47.8050)
    \put(0,0){\includegraphics[width=\textwidth]{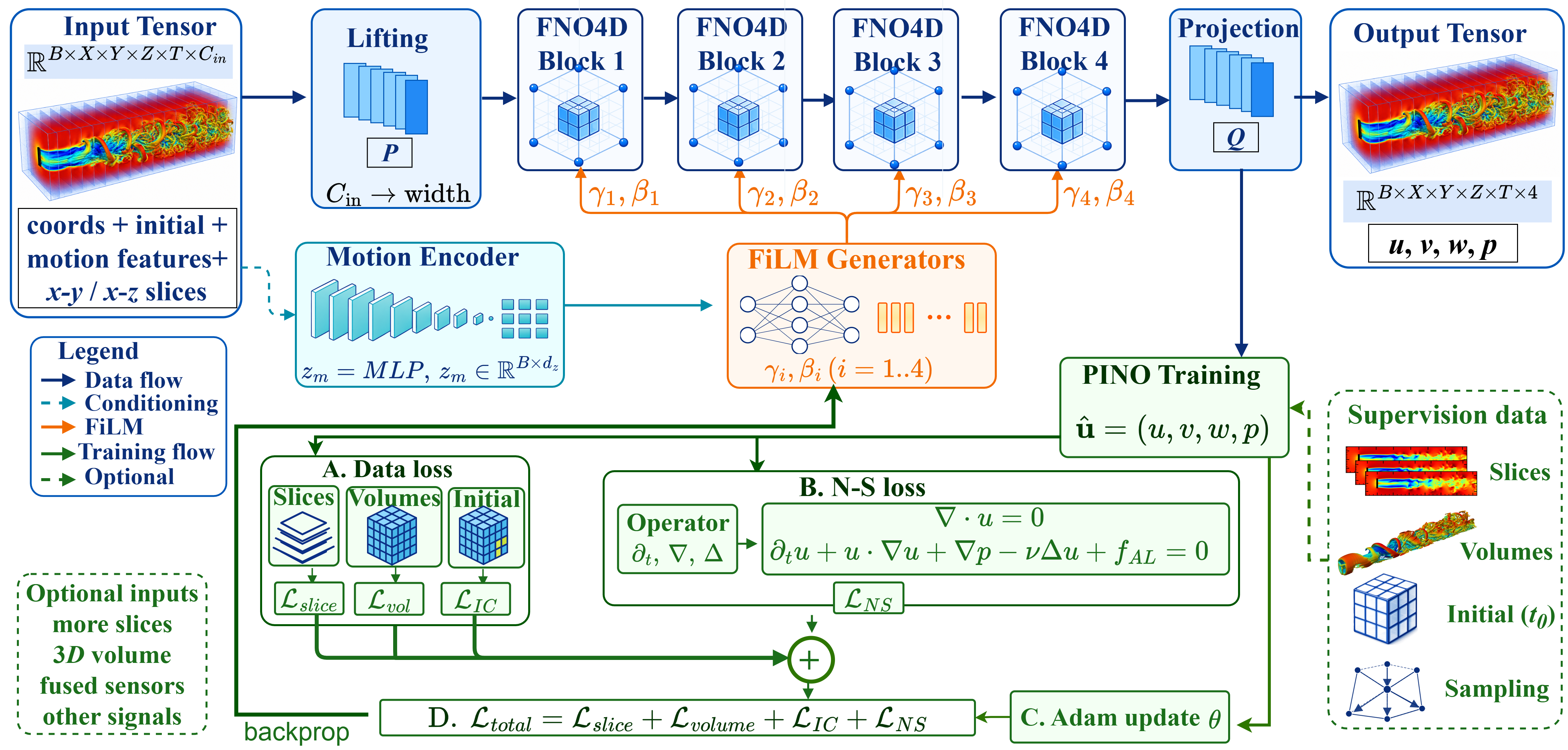}}
    \put(49.10,13.65){\colorbox{diagramgreen}{\makebox(28.8,1.8){\figannotationtext{0.72}{Continuity residual: Eq.~\eqref{eq:div_res}}}}}
    \put(49.10,11.55){\colorbox{diagramgreen}{\makebox(28.8,2.0){\figannotationtext{0.72}{Momentum residual: Eq.~\eqref{eq:mom_res}}}}}
    \put(53.65,8.75){\colorbox{diagramgreen}{\makebox(3.6,1.65){\figannotationtext{0.7}{$\mathcal{L}_{\rm res}$}}}}
    \put(25.20,1.10){\colorbox{white}{\makebox(35.8,1.75){\figannotationtext{0.72}{D. Weighted objective: Sec.~\ref{sec:training_objective}}}}}
  \end{picture}%
  \endgroup
}